\documentclass[fleqn,10pt]{wlscirep}
\usepackage[utf8]{inputenc}
\usepackage[T1]{fontenc}
\usepackage[utf8]{inputenc}
\usepackage[T1]{fontenc}
\usepackage{multirow}
\usepackage{sidecap}
\usepackage{xcolor}
\usepackage{tcolorbox}
\usepackage{comment}
\title{One pathogen does not an epidemic make: A review of interacting contagions, diseases, beliefs, and stories}
\author[1,2,3,4,*]{Laurent H\'ebert-Dufresne}
\author[5]{Yong-Yeol Ahn}
\author[1,6]{Antoine Allard}
\author[7]{Vittoria Colizza}
\author[2,8]{Jessica W. Crothers}
\author[1,3,3]{Peter Sheridan Dodds}
\author[1,4,9]{Mirta Galesic}
\author[10]{Fakhteh Ghanbarnejad}
\author[11]{Dominique Gravel}
\author[4,12,13]{Ross A. Hammond}
\author[14]{Kristina Lerman}
\author[1,3]{Juniper Lovato}
\author[2,15]{John J. Openshaw}
\author[1,4]{S. Redner}
\author[1,4,16]{Samuel V. Scarpino}
\author[17,18]{Guillaume St-Onge}
\author[19]{Timothy R. Tangherlini}
\author[1,2,20]{Jean-Gabriel Young}
\affil[1]{Vermont Complex Systems Institute, University of Vermont, Burlington, Vermont, USA} 
\affil[2]{Translational Global Infectious Disease Research Center, University of Vermont, Burlington, Vermont, USA} 
\affil[3]{Department of Computer Science, University of Vermont, Burlington, Vermont, USA} 
\affil[4]{Santa Fe Institute, Santa Fe, New Mexico, USA} 

\affil[5]{Center for Complex Networks and Systems Research, Luddy School of Informatics, Computing, and Engineering, Indiana University, Bloomington, IN, USA}
\affil[6]{D\'epartement de physique, de g\'enie physique et d'optique, Universit\'e Laval, Qu\'ebec (Qu\'ebec), Canada} 
\affil[7]{Sorbonne Université, INSERM, Pierre Louis Institute of Epidemiology and Public Health,
Paris, France}
\affil[8]{Department of Pathology and Laboratory Medicine, Robert Larner, M.D. College of Medicine, University of Vermont, Burlington, Vermont, USA}
\affil[9]{Complexity Science Hub, Vienna, Austria}
\affil[10]{School of Technology and Architecture, SRH University of Applied Sciences Heidelberg, Leipzig, Germany.} 
\affil[11]{D\'epartement de biologie, Universit\'e de Sherbrooke, Sherbrooke, Qu\'ebec, Canada}
\affil[12]{Public Health, Brown School,
Washington University in St. Louis, St. Louis, MO, USA} 
\affil[13]{Economic Studies, The Brookings Institution, Washington, DC, USA} 
\affil[14]{Information Sciences Institute, University of Southern California}
\affil[15]{Division of Infectious Diseases and Geographic Medicine, Stanford University, Stanford, CA, USA}
\affil[16]{Institute for Experiential AI, Northeastern University, Boston, MA, USA} 
\affil[17]{Laboratory for the Modeling of Biological and Socio-technical Systems, Northeastern University, Boston, MA, USA} 
\affil[18]{The Roux Institute, Northeastern University, Portland, ME, USA} 
\affil[19]{Department of Scandinavian and School of Information, University of California Berkeley, Berkeley, CA, USA}
\affil[20]{Department of Mathematics \& Statistics, University of Vermont, Burlington, Vermont, USA} 
\affil[*]{laurent.hebert-dufresne@uvm.edu}

\begin{abstract}
From pathogens and computer viruses to genes and memes, contagion models have found widespread utility across the natural and social sciences. Despite their success and breadth of adoption, the approach and structure of these models remain surprisingly siloed by field. Given the siloed nature of their development and widespread use, one persistent assumption is that a given contagion can be studied in isolation, independently from what else might be spreading in the population. In reality, countless contagions of biological and social nature interact within hosts (interacting with existing beliefs, or the immune system) and across hosts (interacting in the environment, or affecting transmission mechanisms). Additionally, from a modeling perspective, we know that relaxing these assumptions has profound effects on the physics and translational implications of the models. Here, we review mechanisms for interactions in social and biological contagions, as well as the models and frameworks developed to include these interactions in the study of the contagions. We highlight existing problems related to the inference of interactions and to the scalability of mathematical models and identify promising avenues of future inquiries. In doing so, we highlight the need for interdisciplinary efforts under a unified science of contagions and for removing a common dichotomy between social and biological contagions.
\end{abstract}

\begin{document}

\flushbottom
\maketitle

\section{The science of contagions}

Arguably, no scientific concept encapsulates the human experience as much as contagions. That is in large part because the situations in which people use the word ``contagion'' are so varied, referring to any process in which a property is passed from one or many agents to others. Mathematically, the concept of contagion then inspires different forms of branching processes \cite{harris1963theory}, dynamical systems \cite{anderson1991infectious}, cascade models \cite{zhou2021survey}, and network systems \cite{porter2016dynamical}. These are used to model human genealogies \cite{watson1875probability}, teaching \cite{morone2004knowledge,kiss2010can}, culture and language \cite{labov2007transmission}, viral trends \cite{jin2013epidemiological, weng2013virality}, scientific ideas \cite{bettencourt2006power}, innovation \cite{rogers2003diffusion}, rumours \cite{dietz1967epidemics, jin2013epidemiological}, misinformation \cite{van2022misinformation,lovato2024diverse}, social movements \cite{mcadam1993specifying, oliver1998diffusion}, and obviously infectious diseases \cite{anderson1991infectious}. Although all of these phenomena interact to shape human life, they are unfortunately often studied in isolation, one at a time. Even simple classifications, such as distinguishing biological and social contagions, may create a false dichotomy and obfuscate their complexity. For instance, epidemics are generally shaped by multiple factors drawn from biological sciences (pathogens, genetics, microbiome) to social sciences (information, culture, behaviour) and anywhere in between (nutrition, life history, living environment). All of these factors can affect both the local mechanisms by which a contagion is transmitted and its global, population-level, presentation that we get to observe.

How do contagions spread? For biological contagions like infectious diseases, our intuition tends to be that if a pathogen occurs in every positive case and is not found in negative cases, it is causally responsible for the contagion as it gets transmitted from one individual to the next. Memes, as self-replicating cultural elements, might play a similar role in social contagions \cite{marsden1998memetics}. This intuition is the basis for the 19th-century postulates of Robert Koch’s \cite{rivers1937viruses}, which established both microbiologic methods and the necessary scientific framework of causality. This framework ushered in the golden age of microbial discovery in which many of the pathogens responsible for the worst diseases of humanity (tuberculosis, cholera, anthrax, rabies, and diphtheria) were identified. Although initially useful, the one pathogen, one disease model ceases to accurately describe the way in which pathogens interact not only with each other but also with the host, environment, and societies in which they exist.

Koch's postulates are too simplistic as they imply that every contagion is caused by a single, sufficient, and necessary pathogen, which is contradicted by empirical evidence \cite{byrd2016adapting, pirofski2012q}. For example, many microbiota-related bacterial populations can improve immune defense and prevent infection by intestinal pathogens \cite{buffie2013microbiota}. Similarly, genetic conditions can spread vertically (from parent to offspring) but also interact with pathogens transmitted horizontally (within a generation), such as resistance conferred against malaria by the gene responsible for sickle haemoglobin \cite{williams2005sickle}. Pathogens might therefore be found in negative cases because some other spreading process is interacting with the contagion. Conversely, one can imagine that pathogens can also be found in negative cases because it has yet to interact with some other synergistic contagion; this is the case, for instance, with opportunistic pathogens \cite{brown2012evolution}. We detail documented biological mechanisms of pathogen interactions further in Box 1.

Similar arguments can explain why the modelling of social contagions should also move away from the assumption that one pathogen equals one contagion \cite{lehmann2018complex}. Obviously, no single piece of information, no single meme, and no single idea spreads in a vacuum. 
Different people can adopt the same ideology for different reasons. 
Information, ideas, or behaviours are related to each other and they interact with each other, obfuscating classical notions of causality. 
For example, can a single piece of information cause anyone to develop an anti-vaccination sentiment? Probably not \cite{kata2012anti, klein2024psychology}, especially since the discourse surrounding epidemics can be quite complicated and nuanced \cite{chen2020tracking}.

There is a dire need for unifying the different contagion frameworks and moving beyond the paradigm of ``isolated contagion'' and the classic distinction between ``simple'' and ``complex'' contagions. 
Indeed, modeling of biological contagions often uses a paradigm called ``simple contagions'' which assumes a linear relationship between exposure and transmission, while modelling of social processes often assumes more general ``complex contagions''.
These complex contagions take different nonlinear functional forms to relate exposure to transmission\cite{guilbeault2018complex}, and they can also rely on mechanisms from social diffusion\cite{rogers2003diffusion},  sensing\cite{hammond2007exploring}, adaptation \cite{alshamsi2015beyond}, or learning\cite{galesic2023beyond}.
This conventional approach to modelling biological or social contagions creates a false dichotomy; contagions almost always interact, and these interactions blur the distinction between contagions of different nature (and models thereof). 
Regardless of its nature and transmission mechanisms, a contagion is often shaped by multiple factors that can be biological, ecological, or social. The effects we wish to consider can be direct interactions between different contagious processes (e.g., two infectious diseases or rumors) or indirect interactions through important covariates (e.g., education or norms). These different types of effects are not always easy to distinguish, given that covariates like education, norms, or culture can themselves be considered as contagious at some level. We provide examples of how contagions interact across scales and scientific domains in Box 2.

\begin{tcolorbox}[title= Box 1: Biological interactions within hosts, float, floatplacement=t!]
\label{tcolorbox:biobox}

It is increasingly apparent that interactions between microbes within and across kingdoms play a critical role at the host-pathogen interface. Infection with the hepatitis D virus (HDV), for example, requires co-infection with hepatitis B virus (HBV) due to its need for HBV-derived replicative machinery to complete its viral lifecycle. Co-infection with these two viruses potentiates one another, leading to more severe clinical disease\cite{negro2023hepatitis}. Microbial interactions become even richer in the context of infectious disease syndromes (pneumonia, diarrhea, skin and soft tissue infections), which are often polymicrobial or the result of simultaneous or consecutive co-infection with multiple pathogens as exemplified by post viral bacterial pneumonia following influenza virus infection\cite{rowe2019direct}.

Things grow even more complicated when considering the impact of diverse microbial community structures, the microbiome. Introduction of bacteriophages can directly shape bacterial community structures while commensal bacteria directly and indirectly support or inhibit growth of other pathogens through nutrient competition, iron scavengering, and secretion of antimicrobial molecules (bacteriocins) that directly inhibit the replication of other pathogens leading to antagonistic microbial dynamics in which growth of one organism directly inhibits the growth of another\cite{lian2022bacterial}. These parasitic microbial interactions in which growth of one organism reduces that of another are underscored by diseases such as \textit{Clostridium difficile} infection, which are a result of a dysbiosis or imbalance in the gastrointestinal microbiome\cite{seekatz2022role}.


Host immune dysfunction can also alter pathogen susceptibility, disease severity, and onward transmission dynamics. Within developed regions, the most common cause of immunosuppression is iatrogenic, resulting directly from the use of immunosuppressing medical therapies\cite{martinson2024prevalence,wallace2021prevalence}. Globally however, the most common cause of immunosuppression is malnutrition, with 230 million children considered nutritionally deficient in 2023\cite{morales2023effects}. The impact of immunosuppression on the differential spread of infectious diseases across global populations is perhaps best highlighted by HIV, which directly infects key cells of the immune system, resulting in acquired immunodeficiency syndrome (AIDS). The synergistic impact of HIV infection on the spread of other pathogens, notably \textit{Mycobacteria tuberculosis} and \textit{Treponema pallidum} (causative agent of syphilis), results in regional syndemics. The overlapping prevalence of the HIV/AIDS and tuberculosis is notable due to the direct immunologic effects of HIV infection, which reduces CD4+ T cell mediated control of the intracellular bacterium, \textit{M. tuberculosis}, promoting its spread across populations and distinction as the number one cause of death in HIV-infected individuals\cite{bell2018pathogenesis}. In the case of syphilis, HIV transmission is enhanced by the ulcerative lesions occurring during primary infection with \textit{T. pallidum} and HIV-induced immune suppression in turn enhances syphilitic disease progression and onward transmission of the \textit{T. pallidum} bacterium\cite{fleming1999epidemiological}. The dynamic transkingdom interactions highlighted here illustrate the need to evolve beyond a one pathogen one disease model.
\end{tcolorbox}

\begin{tcolorbox}[title= Box 2: Interacting contagions across domains, float, floatplacement=t!]

Epidemics can be the result of multiple interacting contagions involving both the biological and social realm. The paths by which misinformation and disinformation can lead to worse outcomes are complex and many but include reducing the willingness of an individual to vaccinate, obstructing efforts to contain outbreaks, amplifying political discord, increasing fear and panic, and worsening the misallocation of resources\cite{do2022infodemics}. The interplay between infectious pathogens and social elements leading to the spread of both disease and infodemics is far from new. For example, concern that cow heads might sprout from the sites of inoculation led to vaccine hesitancy, which hindered attempts to control smallpox\cite{jin2024social}. 

More modern examples involving the spread of infectious diseases are also plentiful: many of these involve measles, a disease which spreads so effectively in humans that small drops in a population’s vaccine coverage can lead to dramatic outbreaks. The 2019 measles outbreak in the Philippines, which caused over 30,000 cases, was likely sparked by an increase in anti-vaccination sentiments\cite{dyer2019philippines} fueled in part by the issues with a tetravalent Dengue vaccine which increased the risk of severe illness when infected with Dengue in those who had not had Dengue prior to vaccination\cite{Normillesafety}. Similarly, anti-vaccine sentiment spreading in Samoa after the tragic deaths of two children following nurses administering vaccine with an expired muscle relaxant instead of water\cite{BBC2019} precipitated a drop in MMR vaccination coverage from over 75\% for the first dose in 2017 to about 40\% coverage in 2018, resulting in a measles outbreak that killed over 80 children in 2019\cite{champredon2020curbing}. 

Examples need not be restricted to infectious diseases. For example, multiple stunts and “challenges” have been fueled by memes and social media and can cause a range of adverse health outcomes. Examples include pulmonary damage from consuming spoonfuls of cinnamon in the “Cinnamon Challenge”\cite{grant2013ingesting}, burns caused by applying salt and ice cubes to the skin in the “Salt and Ice Challenge”\cite{breakey2015salt}, and antihistamine overdoses causing deaths in teenagers participating in the “Benadryl Challenge”\cite{FDA2020}.

While these examples showcase interactions between the biologic and the social leading to worse outcomes, this interaction does not need to be negative: messaging, for example, can help people make decisions that lead to better health outcomes and could make epidemics less likely to occur. While studies have attempted to identify components that may lead to successful and productive public health messaging campaigns\cite{merchant2021public, hyland2021toward}, epidemics are rare enough events that it is difficult to truly evaluate and generalize the requirements for a successful campaign.

Importantly, public health messages and data interact with ongoing epidemics and can contribute to the inaccuracy and inconsistency of models. Recent work studies the resulting stochastic dynamics in the fundamental epidemic
reproductive number $R_0$\cite{krapivsky2024epidemic}. If the epidemic is perceived as acute, the
public may readily acquiesce to behavioural restrictions, such as isolating,
masking, vaccinating, etc., so as to reduce the reproductive number. If the epidemic is perceived as decaying, then the public will
want to relax their vigilance.  This latter aspect was reflected in the
extensive and sometimes vitriolic debate about the efficacy, or even the
utility, of various mitigation strategies. This interaction results in a tug-of-war between public-health mandates and social opinions about how to respond to an epidemic, resulting in huge fluctuations in temporal dynamics\cite{krapivsky2024epidemic}.
\end{tcolorbox}

\begin{table}[]
    \centering
\begin{tabular}{ |p{1.8cm}||p{3.4cm}|p{3.2cm}|p{3.3cm}||p{3.3cm}|  }
 \hline
 \textbf{Interaction}& \textbf{Biological example} &\textbf{Social example}& \textbf{Mixed example} & \textbf{Synonyms}\\
 \hline
 Synergistic   & HIV and syphilis\cite{fleming1999epidemiological}    & Related memes/ideas\cite{pathak2010generalized}&   Disease \& anti-vacc.\cite{mehta2020modelling} & Cooperative\\
 Antagonistic&   Bacterial competition\cite{lian2022bacterial}  & Political opinions\cite{sznajd2005left}   & Influenza \& vaccines\cite{fu2017dueling} & Competitive/dueling\\
 Parasitic & \textit{C. difficile} \& commensal enteric bacteria\cite{seekatz2022role} & Fake news \& fact checking\cite{tambuscio2015fact} & Disease \& awareness of the disease\cite{hebert2020spread} & Asymmetric \\
 \hline
\end{tabular}
    \caption{Examples of interacting contagions across domains. Directionality and strength of the association is not always easy to assess. For example, vaccination against influenza might remain strong even after or during a weak flu season, but awareness of a new emerging pathogen will only spread if the pathogen does as well.}
    \label{tab:examples}
\end{table}

The renewed interest in interacting contagions brings more to the table than new keywords like ``syndemics'' for synergistic epidemics and ``infodemics'' for epidemics of (mis-)information \cite{briand2021infodemics}. It also brings a fresh post-disciplinary perspective on the problem and a call for integrated efforts\cite{bedson2021review}. 

In what follows, we do a targeted review of the techniques for modeling interaction contagions in an attempt to highlight some key elements from that literature and map a way forward. In Sec.~\ref{sec:physics}, we first explain how, following lessons learned from other complex systems \cite{anderson1972more,hebert2024path}, ``more contagions are different contagions.'' We discuss how the behaviour of interacting contagions differs from the conventional wisdom built on models of individual contagions. This makes contagion forecasting, an inherently noisy process\cite{krapivsky2024epidemic}, even more intractable in the face of unknown interactions. In Sec.~\ref{sec:social}, we then review novel empirical and data-driven efforts, mostly coming from interacting cascades of social contagions on social media and in the science of stories. In Sec.~\ref{sec:beliefs}, we give one particularly potent example of how contagions can interact within hosts using the study and dynamics of beliefs. Finally, in Sec.~\ref{sec:ecology}, we attempt to outline promising ecological perspectives that might help the science of interacting contagions avoid the trap of high model dimensionality.

\section{The physics of interacting contagions\label{sec:physics}}

Mathematical modelling of contagions has a long history going back at least to the 1700s, with the creative work of Daniel Bernoulli\cite{bernoulli1760essai} among others. In Bernoulli's work, a mathematical system of smallpox dynamics is developed, but already, the text highlights the interaction of the disease with indirect factors such as the age of individuals and the dynamics of inoculation as a method of prevention. Most modern mathematical models borrow a structure and a set of assumptions from a model called Susceptible-Infected-Removed (or susceptible-infectious-recovered, the SIR model) published in 1927 by Kermack \& McKendrick\cite{kermack1927a}. In this model, infected agents transmit the contagion to their susceptible neighbour at a given rate and are removed from the dynamics through death or immunity at some other rate. One can easily generate variants of this model. For example, the SIS process assumes that the disease does not confer long-lasting immunity such that recovered individuals return directly to the susceptible state. The SI process, in turn, assumes that there is no recovery at all.

Early studies of multiple contagions built on the SIR and SIS foundations to couple the dynamics of contagion with the evolutionary dynamics or cross-immunity between biological pathogens\cite{may1995coinfection, andreasen1997dynamics} or the social process of disease awareness\cite{epstein2008coupled}. These new models often assumed that contagions were unaffected by each other \cite{may1995coinfection}, or that one was necessary for the other \cite{newman2013interacting}, or that they were competing for a chance to spread \cite{andreasen1997dynamics, gog2002dynamics, karrer2011competing, bansal2012impact, miller2013cocirculation, poletto2013host, gleeson2014competition, williams2022immunity}, or directly in opposition to each other \cite{newman2005threshold, epstein2008coupled, funk2009spread, funk2010interacting, marceau2011modeling}. These efforts have previously been reviewed in detail \cite{wang2019coevolution}. And while there have been studies on a broader definition of interactions between contagions \cite{vasco2007tracking, abu2008interactions,hebert2013pathogen,campbell2013complex,de2016physics,fu2017dueling, liu2018interactive}, the mathematical modelling community has not quite moved beyond the simplest case of two contagions interacting in simple, often symmetric and deterministic, ways. This is due in part to the richness of the behaviours that emerge even in this simple case, and in part because considering more interacting contagions makes the dimensionality of our models grow exponentially. 

\begin{figure}
    \centering
    \includegraphics[width=0.9\linewidth]{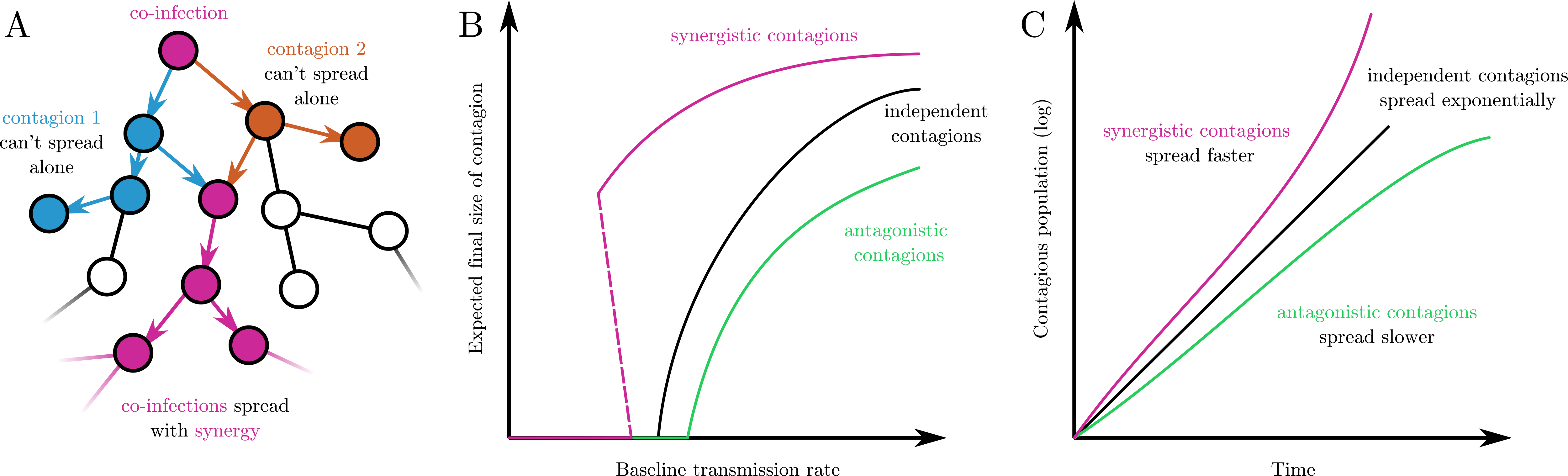}
    \caption{Illustration of the dynamics of interacting contagions. \textbf{A}: Schematic representation of two synergistic contagions (blue and orange) spreading synergistically through a network. Co-infections, shown in pink, are needed to sustain the contagion and are aided by clustering. 
    \textbf{B}: Phase transition of various types of contagions. While independent contagions display continuous transitions, synergistic contagions can build up transmission potential, which leads to discontinuous transitions reminiscent of physical systems where latent heat accumulates. \textbf{C}: Growth rate of synergistic (pink), independent (black), and antagonistic (green) contagions. Synergistic contagions tend to grow super-exponentially since they get more likely to interact as they spread further.}
    \label{fig:physics}
\end{figure}

Importantly, the physics of interacting contagions is more than that of the simple sum of independent contagions, especially when contagions interact synergistically as they spread~\cite{chen2013outbreaks, hebert2015complex, cai2015avalanche, chen2017fundamental, soriano2019markovian, grassberger2016phase, hebert2020macroscopic, lamata2024pathways}. In a simple contagion model, say the SIR model described above, there is a monotonic relationship between the transmission rate and the final size of the contagion \cite{anderson1991infectious}. Interestingly, depending on the density of contacts in the population, there exists a critical value of transmissibility below which the expected size of the contagion is zero, and above which it increases monotonically. But importantly, this phase transition between a contagion-free state and an endemic contagion is typically continuous. That is not the case for synergistic \cite{hebert2015complex} or cooperative \cite{cai2015avalanche} contagions, where two spreading processes can increase each other's transmissibility. These models can assume, for example, that a contact between a contagious agent and a susceptible agent transmits at a fixed rate $\lambda$, or a higher rate $\lambda'>\lambda$ if a second contagion is involved. This effect might be the same if the second contagion affects the contagious agent (e.g., a superspreader) or the susceptible agent (e.g., increased susceptibility). Regardless of the details, the system can get into a frustrated state where many contagious agents can only transmit the contagion to their susceptible neighbours if a second contagion also reaches them (see Fig.~\ref{fig:physics}A). These frustrated transmissions are similar to latent heat in physical systems, leading to discontinuous phase transitions. Or, in this case, the discontinuous emergence of a large contagion as we tune the transmission rate (see Fig.~\ref{fig:physics}B).

There are other elements of conventional wisdom from simple independent contagions that do not transfer to the dynamics of interacting contagions. Features of the structure of the underlying contact network that tend to slow down contagions, like triangles or any form of clustering, can now hasten the spread of synergistic contagions~\cite{hebert2015complex}. The intuition behind this phenomenon is simple. For an independent contagion, the optimal structure on which to spread is a treelike network where it is impossible to backtrack when starting from a single patient zero. In that case, all connections lead to new, and therefore susceptible, agents. Clustering of connections can instead trap a contagion and cause transmission efforts to be ``wasted'' on already contagious agents. However, synergistic contagions benefit from being kept together. On a treelike network, a co-infected patient zero might transmit one contagion to one branch of the tree and another to a different branch. These contagions might then not interact with one another for a long time and, therefore, not benefit from their synergy. A small amount of clustering helps them stay together. 

A similar effect is found in parasitic contagions, where one contagion benefits from the presence of another contagion but hinders its transmission in return. This is the case, for example, of positive messaging around a negative contagion, like the spread of preventive awareness against an infectious disease. The awareness is more likely to spread when the disease is spreading, and individuals exposed to the disease are more likely to be receptive to relevant information. Conversely, the preventive message is likely to reduce the transmission risk of the disease. The awareness is thus a parasitic contagion to the epidemic \cite{hebert2020spread}. With asymmetric interactions, clustering can hurt awareness more than the disease, leading again to a larger epidemic than expected without clustering.

Finally, synergistic contagions can also spread superexponentially since the more contagions spread, the more likely they are to interact and benefit from their synergy, thus further accelerating their spread (see Fig.~\ref{fig:physics}C). This dynamic clashes with most mechanistic prediction models~\cite{hebert2020macroscopic}. As contagions spread and the number of contagious agents grows, the more likely co-infections become such that positive interactions become more frequent and important. For a time series of either contagion, this simple statistical effect leads to what looks like an accelerating spread or a transmission rate that increases with time.  Altogether, basic intuition built by focusing on a single pathogen might not hold when looking at an ecology of interacting contagions. New modelling approaches are needed. Thankfully, ideas from the social sciences might again come into play. 
 
All of the features described above are also signatures of complex contagions, as introduced in the previous section. In complex contagions, also called frequency-dependent models, transmission depends nonlinearly on the number or frequency of exposures such that the growth rate of a contagion can vary as it spreads \cite{rogers2003diffusion, guilbeault2018complex}.
Recent results show that synergistic contagions can be mathematically indistinguishable from these nonlinear effects in complex contagions, such as the psychological phenomenon of social reinforcement~\cite{hebert2020macroscopic}. Like peer pressure, this phenomenon means that a second exposure to an idea or behaviour through a second source is more effective than the first exposure~\cite{centola2010spread, guilbeault2018complex}. For example, ten friends telling you to read an article are more likely to convince you than one friend telling you the same thing ten times. That is usually not true for biological contagions, where our models assume a linear relationship between infection rate and exposure regardless of the source. Indeed, researchers often use simple linear contagions for biological applications and complex nonlinear contagions with reinforcement for social applications. Studies suggest the distinction breaks down once contagions interact.

In recent years, modelling studies have also broadened the concept of interaction structure and focused on networks with different (multilayer) interactions \cite{de2018fundamentals} or with group (higher-order) interactions\cite{wang2024epidemic}. There has therefore been a rapid influx of studies generalizing these previous models. Given that nonlinear group interactions alone create apparent complex contagions, the phenomenology of higher-order and interacting contagions is only richer. Studies have looked at competing contagions in multilayer networks where different contagions spread on different sets of contact\cite{funk2010interacting,marceau2011modeling,brodka2020interacting, huang2021modeling}, through higher-order dynamics\cite{li2022competing, nie2022markovian, li2023coevolution}, and even both \cite{fan2022epidemics, you2023impact, liu2023epidemic, hong2023coupled, zhu2023epidemic}. These extensions are natural, and in fact, some of the early studies on cooperating epidemics also assumed a multilayer structure \cite{min2014diversity} or a group structure \cite{hebert2015complex}.

The rich dynamics of interacting contagions pose an important challenge: Is it possible to measure contagion mechanisms without explicitly controlling for all possible interactions and covariates? Social reinforcement can be defined and measured by the increase in infection probability caused by repeated exposures compared to a null model of equivalent exposures. However, this measurement might be impossible to distinguish from interactions among pathogens. Did someone share a meme because many of their friends liked it, or had they previously shared a related meme? The flip side of this challenge is that measuring social reinforcement can then be used as a proxy for interactions between contagions. Recent work has taken this perspective to justify the need for tractable nonlinear models \cite{nguyen2024upper}. New effective models for interacting contagions are needed, especially since we know that countless contagions of biological and social nature interact at all times and that the dimensionality of our current models grows exponentially with the number of contagions involved.

\section{Modeling social contagions and their interactions\label{sec:social}}

Attempts to operationalize and model social contagion processes first appeared in the second half of the 20th century. In the 1950s, Katz and Lazarsfeld\cite{katz1955a} published ``Personal Influence''\cite{katz1955a}, a book which would become highly influential about the nature of influence. Their finding were based on their in-the-field study of decision making by women in the city of Decatur, Illinois. Their major contribution was to introduce the ``two-step'' model of influence, whereby media reached local opinion leaders who then influenced their friends and acquaintances. Though Katz and Lazarfeld were clear that their two-step model was a non-universal approximation of true social influence, the idea that there exists a class of special, influential people---and consequently the importance of identifying and engaging them for means of large-scale social change to whatever ends---would become a hardened, enduring, and broadly popularized concept\cite{weimann1994a,rogers2003diffusion,gladwell2000a}. As the real-data-informed field of network science developed in the late 1990s, the notion of opinion leaders would be challenged with more sophisticated social network models\cite{watts2007a}. Nevertheless, the two-step model moved the understanding of media influence beyond the traditional ``hypodermic model,'' revealing and elevating the role of social networks in shaping collective and individual opinions and beliefs.

In the 1960s, early considerations of how to mathematically operationalise social contagion fell to simply porting across the SIR model of mathematical epidemiology\cite{kermack1927a}. In a 1964 Nature article, Goffman and Newill explicitly cast the spread of an idea as akin to the spread of a disease\cite{goffman1964a}. It is worth reflecting on the scope and strength of their framing:
\begin{quote}
  ``For example, consider the development of psychoanalysis in the early part of this century. Freud was no less host to the infectious material of the `disease' of psychoanalysis than the person carrying the organism capable of transmitting a cold, nor is his writing less of a `vector' carrying the `infectious material' than the mosquito as a carrier of malaria. Jung might represent an example of acquired resistance to the disease while the resistance of the medical community of Vienna could represent innate immunity. The development of the psychoanalytic movement \ldots was in its way no less an `epidemic' than the outbreak of influenza in 1917 and 1918.
  
  One can argue similarly that Darwin and evolution, Cantor and set theory,  Newton and mechanics, and so on,   were examples of `epidemics' in the world of scientific thought which were instigated by the introduction of a single infective into a population. The analogy is not restricted to science; for examples such as Christ, Buddha, Moses and Mohammed can be cited in the religious field \ldots''
\end{quote}

Of course, social contagions are inherently different from biological contagions as there are no ``cultural pathogens'' easily identified and cultured\cite{cavalli1982theory}. Instead, as recent studies suggest, social contagions are always shaped by other effects such as homophily or cultural\cite{ananthasubramaniam2024networks}, human biases and demographic heterogeneity\cite{lovato2024diverse} or because of dislike and distrust between subpopulations\cite{lerman2024affective}.

Clear movement beyond SIR models began (at least) in the early 1970s, when Schelling introduced  his physical, checkerboard-based model of self-organizing neighbourhood segregation processes\cite{schelling1971a}. Schelling operationalized social contagion with the concept of thresholds: Individuals adopt (or reject) a characteristic (action, belief, behaviour, etc.) based on the fraction of those around them with that characteristic or some other influential attribute. While simple to run, Schelling's model was non-trivial to address analytically. In 1978, Granovetter showed that absent any interaction structure, a mean-field threshold model produced informative stories of social contagion\cite{granovetter1978a,granovetter1983a,granovetter1986a,granovetter1988a}. Like Schelling's model, a seemingly moderate population of could universally adopt a behaviour. These models showed mechanistically how collective uniformity could arise not from individual uniformity but rather social following.

It is worth noting that while seemingly distinct, the essential models of biological and social contagion can be reconciled. For example, a generalized contagion model incorporating memory successfully interpolates between SIR-type models and threshold models\cite{dodds2004a,dodds2005a}.

Abstract modelling of contagions, regardless of their nature, can only get us so far. Turning to data to validate these theories in specific domains is critical. A major, enduring problem with all empirical social network contagion is missing data, either through limited sampling of a social network (hence missing links and interactions), or due to interactions occurring outside of the sphere of observation (e.g., through direct text messages between individuals, or influence from other media across scales).

Modern social media platforms can control and monitor what people see and how they behave with high precision. But the algorithms behind major platforms are proprietary and hidden. Observational analyses of online behaviour and properly formalized academic experiments promise an ethical way forward. Because of the massive data on offer,  social media provides a unique lab bench to both characterize social contagion and test theoretical models.

As a first simple proxy for social contagion, one can look at cascades of re-sharing of the same content on social media (resharing a post, keyword, or a given URL). Under that lens, while co-infection data regarding biological contagions can be rare, social media are essentially a messy soup of countless interaction contagions\cite{guille2013information} shaped by each other and by the social networks that support them\cite{lerman2010information}. Therefore, it is not surprising that online social networks inspired some early studies of multiple spreading processes\cite{pathak2010generalized}. These processes can interact by competing or complementing each other\cite{lu2015competition}, and telling these mechanisms apart from noisy data can be subtle but is not impossible\cite{myers2012clash,zarezade2017correlated}. For example, Zarezade et al. \cite{zarezade2017correlated} highlight that incorporating interactions into models enhances their predictive accuracy of social media cascades, even when accounting for the added model parameters. In their model, a contagion spreads to susceptible individuals at a rate specified by a Hawkes process around each susceptible individual\cite{hawkes1971spectra}, a self-exciting process that relates the transmission rate of a contagion to the sum of recent exposures to the contagion itself \textit{and} to related contagions. Related work focused on competition between memes obtained similar results, showing that the competition for potential ``hosts'' and their attention is sufficient to qualitatively explain the broad diversity in popularity, lifetime, and activity of memes\cite{weng2012competition}. In that model, competition between memes occurs through a mechanism of limited attention, through which a susceptible agent is only really exposed to a finite amount of memes in its recent memory. The general idea being that signatures of online behaviour are a reflection of recent exposure to contagions both online or in the real world, and both social and biological. In fact, a whole subfield of digital disease surveillance has also looked at the potential of using social media data as a proxy to track the spread of infectious diseases through keywords, posts, and searches related to a disease or its symptoms\cite{althouse2015enhancing}.

Despite this abundance of data, monitoring of social contagions is complicated by the fact that the signature of any contagion can mutate quickly\cite{leskovec2009memetracking}. Early methods often focused on tracking specific hyperlinks or proper names\cite{gruhl2004information, adar2005tracking}, which is appropriate for short timescale before new hyperlinks or names start describing the same information. Other approaches tend to aggregate many cascades, such as mixture models or dynamic topic models\cite{kleinberg2002bursty}, which by analogy would be similar to tracking respiratory illnesses without tracking individual virus families\cite{mandl2004implementing}. To follow a social contagion as it evolves, it is possible to track specific short phrases that are unlikely to mutate\cite{leskovec2009memetracking}, but this approach is hard to generalise to visual non-text memes. More recent work has therefore moved to multi-modal deep learning models to identify memes\cite{dubey2018memesequencer}, classify them in families\cite{afridi2021multimodal}, and attempt to understand their relations\cite{beskow2020evolution} or predict their potential for virality\cite{ling2021dissecting}.

Beyond the spreading of internet memes and the modeling of social contagions lies a developing, data-driven science of stories and beliefs. That stories matter profoundly to people and societies seem to be both unclear and obvious: Stories are portrayed by some as being just for entertainment, while by others as the core of being human\cite{sartre1938a,ranke1967a,fisher1984a,gould1994a,ferrand2001a}. Even so, the centrality of stories has gained ground, as has the conception and possibility of measuring stories through distant reading\cite{bruner2003a,boyd2010a,moretti2013a,dodds2013b,gottschall2013a,maruna2015a,shiller2017a,puchner2017a,shiller2017a}. Understanding how stories develop, spread, interact, and compete with each other is of utmost importance to understanding social phenomena; from myths and conspiracy theories to hate speech and counter speech\cite{garland2022impact}. Operationalizing the measurement of stories and interactions between story elements is critical to advancing these aims (see Box 3).

\begin{tcolorbox}[title= Box 3: Narrative interactions within and across stories, float, floatplacement=t!]
\label{tcolorbox:storybox}


In crisis situations, there is a strong predilection to construct narratives in which some outside force threatens the integrity of an inside group~\cite{labov1997narrative, barkun2013culture}. 
These narratives allow people to collaborate in shaping the story, resulting in a consensus on who is ``inside,'' who is ``outside'' (and therefore can pose a threat), the range of potential threats, and the possible responses to those threats~\cite{holur2022side,holur2023my}. 
Evaluative comments, either as framing devices or within the narratives themselves, offer clues to the stance of storytellers and their audiences about the acuity and severity of the threat and endorsement of particular strategies or outcomes~\cite{tangherlini2016mommy}. 

Stories of this nature---and the conversations in which they are embedded---can be modeled as an interactant narrative framework graph, where the nodes consist of actants (individuals, groups, institutions, places) and the edges consist of the relationships between those actants~\cite{greimas1966elements, lehnert1981plot}. The graph, extracted from conversations, models the range of possible actants and their interactions; and a story, either complete or in part, can be modeled as a directed acyclic graph activating some subset of the nodes and edges~\cite{tangherlini2020automated}.  

As people react to the unfolding crisis and its attendant threats and disruptors, new actants and relationships are proposed (e.g., by new posts to the forum) and possibly added into the graph, while others are less frequently activated and disappear. Several mechanisms limit the graph from growing too large, including the concept of “tradition dominants,” so that new actants who duplicate the role of existing actants in the graph are subsumed into the existing actant’s role, and the “law of self-correction” so that variations on assignations to insider/outsider status, the range of threats and possible reactions are aligned with existing conditions~\cite{boole1854investigation,eskerod1947lrets,anderson1923kaiser}. These mechanisms and constraints allow new stories to contribute to the evolving conversation, while the constraints on the graph provide support for the group’s beliefs and narrative motivation for actions based on those beliefs. Consequently, the storytelling environment not only reacts to changes in external conditions but effects those external conditions through real-world actions while maintaining a degree of narrative stability.

    \begin{minipage}[t]{\linewidth}
    \centering
    \vspace*{0pt}
        \includegraphics[width=0.5\linewidth]{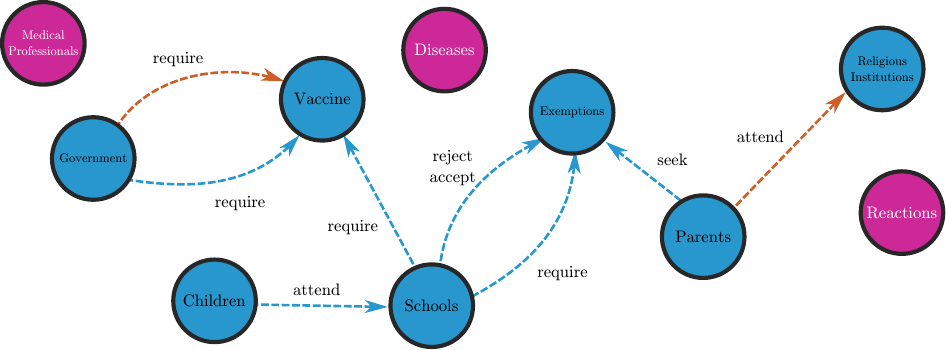}
        \captionof{figure}{A post about religious exemption to vaccination requirements, ``A friend of my daughter got her `shots' for this one and now has multiple health issues. Anxiety attacks, rashes, and a lot of fatigue. My girl is not getting them. No way\ldots'' can be represented as an interaction subgraph between concepts of the broader discussion forum, which comprises thousands of interlocking stories and story parts.}\label{fig:boxfigure}
    \end{minipage}
\end{tcolorbox}

\section{The spread of beliefs as interacting contagions\label{sec:beliefs}}

\begin{SCfigure}[][h] 
\includegraphics[width=0.4\linewidth]{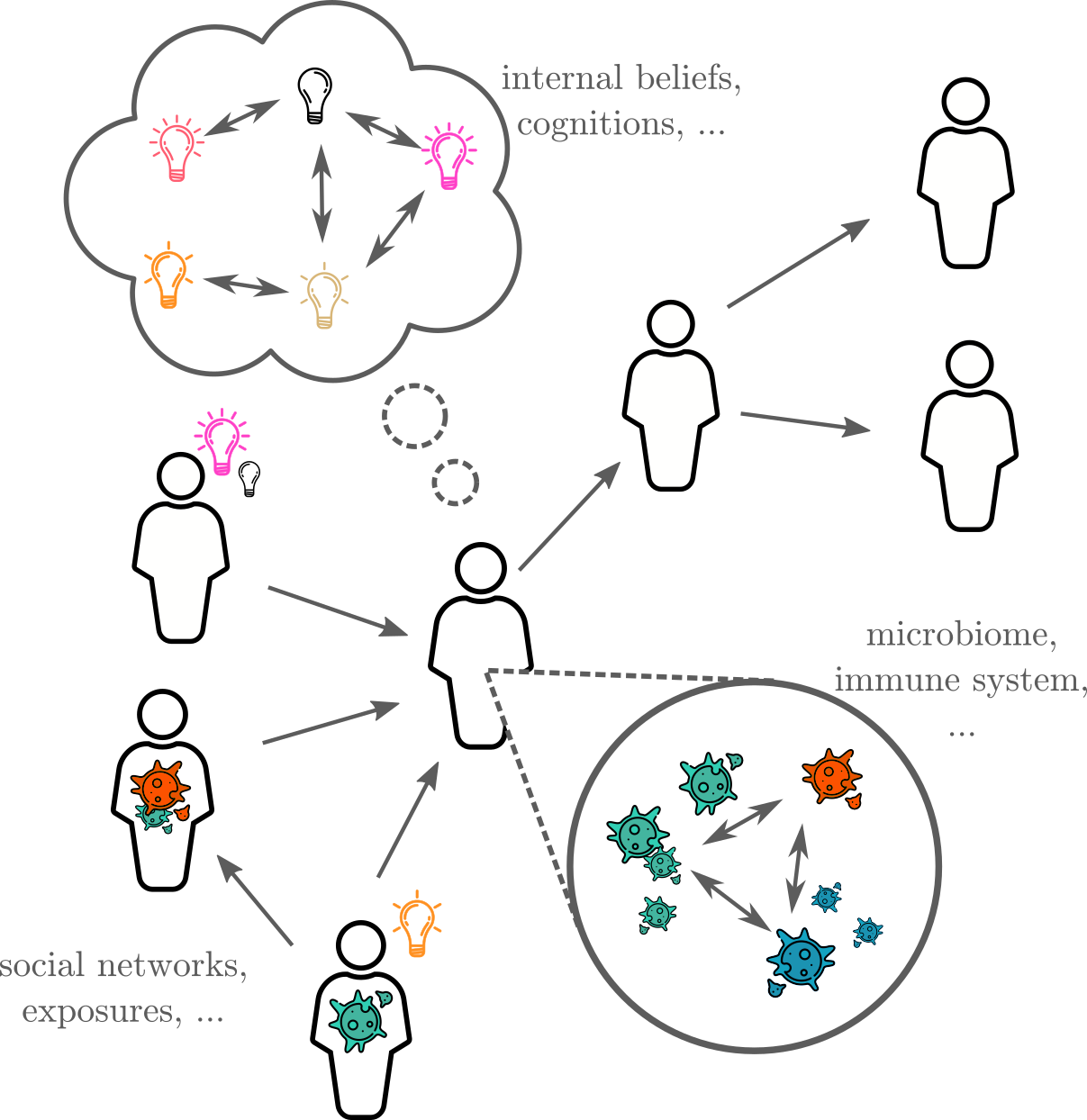}
\caption{Interactions across domains and scales. Biological and social contagions (here represented by coloured viruses and different light-bulbs, respectively) spread through contact networks and shared physical or digital environments. Contagions can interact directly in the environment or during contacts. When an agent is exposed to new viruses or ideas, the potential transmission is mediated through other layers of interactions, now internal. New beliefs or ideas interact with an existing belief system. Likewise, new pathogens interact with an existing microbiome and immune system. These multi-layered interactions can all drive the observed dynamics of contagions \cite{bauch2013social, magarey2020information}.}
\end{SCfigure}

Interacting contagions in social networks also differ from biological contagions as they often spread in signed social networks that explicitly distinguish between positive links (representing trust and favoritism, generally toward in-group members) and negative links (representing distrust and enmity, generally toward out-group members). Positive links facilitate the spread of contagions within the group of like-minded individuals, while negative links can lead to rejection or counter-adoption of ideas from distrusted sources. This sign-based framework provides essential structure for understanding how contagions compete across partisan divides. Recent dynamical modeling~\cite{nettasinghe2024dynamics } reveals that the relative strength of positive versus negative links determines whether the social network converges toward consensus or diverges toward polarization, explaining the rapid polarization of stance toward masking and lockdowns during the COVID-19 pandemic~\cite{nettasinghe2024group}.

While much of the work we have surveyed so far emphasizes the spread of contagions across agents, beliefs rarely spread in isolation from other beliefs and related cognitions such as knowledge, social norms, and emotions\cite{ajzen1991theory,monroe2008general}. Therefore, beliefs and opinions (which we here use somewhat interchangeably\cite{olsson2024analogies}) spread not only across social networks but also within internal belief systems.
This interaction of contagions across levels makes the study of beliefs a prominent and powerful example of interacting contagions. 

In social networks, the spread of some beliefs is often easier after a group has already been ``infected'' with a related set of beliefs. During the COVID-19 pandemic, people who grew up in the former East Germany and were thus more accustomed to the idea that the government can enforce certain behaviours, were more likely to support the idea of mandatory vaccination \cite{schmelz2021overcoming}. Beliefs about whether abortion should be banned or not are strongly related to prior 'infections' of a group by particular political and religious beliefs \cite{pew_abortion}. Beliefs about what is normal and desired in a group can change in line with shifting perceptions of what the majority of group members believe or do \cite{festinger1954theory}, with examples ranging from the support for gay rights \cite{gallup_marriage} to the support for extreme political views \cite{ebner_extremism}.

Beliefs also spread within individual minds, affecting existing beliefs and related cognitions and clearing the path for 'invasions' of a belief system with other novel beliefs. For example, when one becomes skeptical about the safety of vaccinations, this can open doors to the development of skepticism for those supporting vaccination, such as scientists and the government. This distrust may make one more likely to accept further related beliefs and conspiracies \cite{pertwee2022epidemic}. Emotions can further facilitate belief spread within an individual's mind. Even a temporary ``infection'' with fear of death can lead one to adopt dislike and prejudice towards beliefs and groups different than one's own \cite{solomon2000pride}.

Models of belief dynamics could profit from incorporating the effects of interacting socio-cognitive contagions. 
Many existing belief dynamics models focus on the spread of only one belief at a time \cite{castellano2009statistical,flache2017models,redner2019reality,olsson2024analogies}. To model the spread of several related beliefs or cognitions more broadly, one can proceed in at least two ways \cite{galesic2021integrating}. 

One way to model the dynamics of several beliefs is to assume that each one is affected by a summary of all the others. Many plausible summary measures have been proposed. Some are normative, such as Bayesian reasoning\cite{chater2006probabilistic,bikhchandani2010theory} or logic \cite{van2011logical}. Others are more descriptive, aiming to mimic actual cognitive processes. Examples are averaging strategies \cite{friedkin1990social}, frequency-based strategies such as plurality or tallying \cite{einhorn1975unit}, birth-death dynamics like Moran processes\cite{fic2024catalysing}, or various non-compensatory strategies such as deciding based on the most important belief or consideration \cite{gigerenzer1996reasoning}. 

Another way to model interacting belief contagions is to model the whole network of beliefs. This idea is not new \cite{converse1964}, but formal models of such networks that can enable analyses of interacting contagions within belief systems have started to be developed only recently. In these network models, nodes are typically beliefs, and edges represent influence between them\cite{dalege2016toward,boutyline2017belief, brandt2019central, vlasceanu2024network}. However, nodes can also represent concepts, and edges represent beliefs about the relationship between them \cite{rodriguez2016collective}. Empirical work has been done to assess the assumptions of these models and their predictions, but this is still a developing area of research\cite{brandt2023inter,turner2022belief, dalege2024networks}.


Modeling interactions as a network of contagions may provide a unified way to explore a wide spectrum of contagion dynamics as emergent behaviours. For example, models of interacting beliefs have been shown to break the dichotomy between simple and complex contagions and integrate both dynamics under a single framework \cite{aiyappa2024emergence}. Linear, simple contagion dynamics occur when the existing belief system is primed to accept a new belief, while complex contagion dynamics occur when a new belief challenges the coherence of the existing belief system, echoing the interacting spreading processes discussed in Sec. \ref{sec:physics}. 


Recent work has coupled the dynamics of disease spread with a spectrum of internal states that can represent awareness or behaviour and can vary according to social contacts through some social sensing process \cite{maghool2019coevolution, ogara2024}. Incorporating the richness of internal belief interactions is a promising direction for a further study of the multiple interactions that occur between and within individuals during social and biological contagions.

\section{The ecology of interacting contagions\label{sec:ecology}}

Across biological contagions, abstract models, stories, and beliefs, the previous sections have hinted at two important challenges in the modelling of interacting contagions. First, the number of states that agents can take grows exponentially with the number of contagions spreading in a population. Therefore, the dimensionality of our models also tends to grow exponentially with the number of contagions we wish to consider \cite{stanoev2014modeling}. Second, contagions can interact on multiple scales. Interactions can occur internally, within agents, mediated by an agent's immune system (biological contagions) or cognition (social contagions). Interactions can also occur across agents, mediated by the environment, transmission pathways, or local culture and norms.

Despite these challenges, there is an obvious need to consider interactions between large numbers of contagions, in the thousands or more, that shape our everyday lives. Therefore, we need to be able to go beyond individual-based models whose dimensionality grows exponentially and start thinking in terms of an ecosystem of contagions. This call for an ecology of contagion is a recent development \cite{guilbeault2018complex}, and we here outline some promising directions. 

There already exist models to infer interactions in large numbers of contagions from detailed time series of social data \cite{myers2012clash} and biological data \cite{shrestha2011statistical}. More rarely, we also sometimes have access to co-infection data, often static counts of individuals infected or involved in multiple contagions. From co-infection data, previous studies have built inference frameworks based on permutation tests\cite{colgate2023network} or joint-species distribution models\cite{fountain2019endemic}. 

Biologically, emerging technologies such as multiplex PCR panels, high-throughput sequencing (metagenomics, metabolomics, single cell sequencing), and organoid model systems, can allow us to begin to better understand the rich interrelated dynamics occurring across microbial communities at pathogen-host interfaces. Integrating these data sets and their insights into contagion models can enhance their accurate reflection of the true biology at play. In the social sciences, it is also a challenge to build broad datasets that include multiple beliefs in a time-resolved manner. Large language models are providing new methods for stance detections and belief quantification that could also resolve interrelated social dynamics\cite{gul2024stance}.

That being said, co-occurrence alone is not evidence of interactions \cite{blanchet2020co}, regardless of how many covariates are included in the analyses. The ideal dataset would, of course, be a time series of individual co-infections, where these existing methods could be combined. Despite the scarcity of such data, a significant number of approaches have already been developed\cite{powell2024systematic}. A more spatial and ecological understanding of co-infections could also help improve inference. This question has been tackled in models\cite{stanoev2014modeling,chen2019persistent}, but separating spatial correlations from actual contagion is not a simple task\cite{waring2023operationalizing}. For social contagions, surveys can help sort these different effects\cite{cavalli1982theory}. For biological contagions, the majority of studies of co-infections rely on observational data or case notes, but a minority of studies attempt to survey specific populations\cite{griffiths2011nature}.

On the modelling side, it is unclear what the ecology of contagions should look like. Can models of food web stability inspire a new generation of models of contagions? Recent models attempt to study the endogenous emergence of interactions between contagions using a game-theoretic perspective\cite{su2016understanding}, evolutionary models\cite{ghanbarnejad2022emergence}, and co-evolutionary interactions with measures to promote or hinder these contagions\cite{khazaee2022effects}. These models encourage us to think of interactions as endogenous or emergent features of contagions and not just as fixed parameters or model mechanisms.

The field of ecology itself is challenged by modelling interacting contagions and currently faces significant analytical limitations related to the issues reviewed above. Levins' metapopulation models, for instance are, very similar to standard epidemiological models \cite{levins1969some}, representing the dynamics of individual patches that could either be empty (susceptible) or occupied (infected) by a given species, and studied for the conditions allowing species persistence at the regional level (outbreaks). Such models have been extended to multiple interacting species \cite{hanski1983coexistence}, but ecologists lost track of the rapidly growing number of potential community states \cite{cazelles2016integration}. 

Solving this modelling challenge may require a fundamental shift in state variables; instead of representing the dynamics of a set of individuals, which could either be susceptible or infected, perhaps a solution would be to represent the dynamics of the growth rates themselves as functions of the entire community of contagious agents. Focusing directly on the contagions rather than on their hosts might help relax assumptions about the binary nature of contagious states or about the different relative timescales of biological, social, and evolutionary contagious forces. With a proper formulation of the dynamical functions, the extensive toolbox of community ecology models could then be used to investigate problems of coexistence, feasibility, stability, and higher-level interactions among contagions.

We thus join our voices to the many recommending ecological models of infectious diseases\cite{hassell2021towards} and contagions\cite{guilbeault2018complex}. This perspective is necessary to appreciate the intricate and dynamic web of interactions between viruses, animals, parasites, humans, behaviours, and beliefs. Furthermore, under this lens, the study of emerging beliefs, stories, or epidemics can then borrow from known theories regarding invasive species in classic ecological modelling\cite{vila2021viewing}.

We end with a question not often posed in the context of these increasingly detailed frameworks. To what end do we pursue these models? The ecological and holistic approach argued above is an increasingly common one, as different fields attempt to produce spatial and population-level understanding in times where social, ecological, and biological variables are in constant flux\cite{lasky2020processes}. Is our goal with these frameworks to actually predict and forecast? Is that truly a litmus test of our understanding of these rich, stochastic, and inherently noisy processes? Even then, what concrete observables are we aiming to predict? New cases of a disease? New believers in a conspiracy theory? Emergence of new contagions or stories? If we adopt an ecological perspective, our goal should probably be to better understand the structure of interactions, and to predict their impact on the stability and hierarchy of existing contagions. 
Importantly, this new objective aims for an ecological science of interacting contagions where we can study contagions as a system in and of themselves, and not just through their individual parts of pathogens and hosts.

\section*{Acknowledgments}
This work is based on the workshop ``Dynamics of interacting contagions'' held at the Santa Fe Institute. The authors acknowledge financial support from the Santa Fe Institute, the Vermont Complex Systems Institute (award \#2242829 from the National Science Foundation), and from the Translational Global Infectious Diseases Research Center of Biomedical Research Excellence (award P20GM125498 from the National Institute of General Medical Sciences).


\begin{thebibliography}{100}
\urlstyle{rm}
\expandafter\ifx\csname url\endcsname\relax
  \def\url#1{\texttt{#1}}\fi
\expandafter\ifx\csname urlprefix\endcsname\relax\def\urlprefix{URL }\fi
\expandafter\ifx\csname doiprefix\endcsname\relax\def\doiprefix{DOI: }\fi
\providecommand{\bibinfo}[2]{#2}
\providecommand{\eprint}[2][]{\url{#2}}

\bibitem{harris1963theory}
\bibinfo{author}{Harris, T.~E.} \emph{et~al.}
\newblock \emph{\bibinfo{title}{The theory of branching processes}},
  vol.~\bibinfo{volume}{6} (\bibinfo{publisher}{Springer Berlin},
  \bibinfo{year}{1963}).

\bibitem{anderson1991infectious}
\bibinfo{author}{Anderson, R.~M.}, \bibinfo{author}{May, R.~M.} \&
  \bibinfo{author}{Anderson, B.}
\newblock \emph{\bibinfo{title}{{Infectious Diseases of Humans: Dynamics and
  Control}}}, vol.~\bibinfo{volume}{28} (\bibinfo{publisher}{Wiley Online
  Library}, \bibinfo{year}{1991}).

\bibitem{zhou2021survey}
\bibinfo{author}{Zhou, F.}, \bibinfo{author}{Xu, X.},
  \bibinfo{author}{Trajcevski, G.} \& \bibinfo{author}{Zhang, K.}
\newblock \bibinfo{journal}{\bibinfo{title}{A survey of information cascade
  analysis: Models, predictions, and recent advances}}.
\newblock {\emph{\JournalTitle{ACM Computing Surveys (CSUR)}}}
  \textbf{\bibinfo{volume}{54}}, \bibinfo{pages}{1--36} (\bibinfo{year}{2021}).

\bibitem{porter2016dynamical}
\bibinfo{author}{Porter, M.~A.} \& \bibinfo{author}{Gleeson, J.~P.}
\newblock \bibinfo{journal}{\bibinfo{title}{Dynamical systems on networks}}.
\newblock {\emph{\JournalTitle{Frontiers in Applied Dynamical Systems: Reviews
  and Tutorials}}} \textbf{\bibinfo{volume}{4}}, \bibinfo{pages}{29}
  (\bibinfo{year}{2016}).

\bibitem{watson1875probability}
\bibinfo{author}{Watson, H.~W.} \& \bibinfo{author}{Galton, F.}
\newblock \bibinfo{journal}{\bibinfo{title}{On the probability of the
  extinction of families}}.
\newblock {\emph{\JournalTitle{The Journal of the Anthropological Institute of
  Great Britain and Ireland}}} \textbf{\bibinfo{volume}{4}},
  \bibinfo{pages}{138--144} (\bibinfo{year}{1875}).

\bibitem{morone2004knowledge}
\bibinfo{author}{Morone, P.} \& \bibinfo{author}{Taylor, R.}
\newblock \bibinfo{journal}{\bibinfo{title}{Knowledge diffusion dynamics and
  network properties of face-to-face interactions}}.
\newblock {\emph{\JournalTitle{Journal of Evolutionary Economics}}}
  \textbf{\bibinfo{volume}{14}}, \bibinfo{pages}{327--351}
  (\bibinfo{year}{2004}).

\bibitem{kiss2010can}
\bibinfo{author}{Kiss, I.~Z.}, \bibinfo{author}{Broom, M.},
  \bibinfo{author}{Craze, P.~G.} \& \bibinfo{author}{Rafols, I.}
\newblock \bibinfo{journal}{\bibinfo{title}{Can epidemic models describe the
  diffusion of topics across disciplines?}}
\newblock {\emph{\JournalTitle{Journal of Informetrics}}}
  \textbf{\bibinfo{volume}{4}}, \bibinfo{pages}{74--82} (\bibinfo{year}{2010}).

\bibitem{labov2007transmission}
\bibinfo{author}{Labov, W.}
\newblock \bibinfo{journal}{\bibinfo{title}{Transmission and diffusion}}.
\newblock {\emph{\JournalTitle{Language}}} \textbf{\bibinfo{volume}{83}},
  \bibinfo{pages}{344--387} (\bibinfo{year}{2007}).

\bibitem{jin2013epidemiological}
\bibinfo{author}{Jin, F.}, \bibinfo{author}{Dougherty, E.},
  \bibinfo{author}{Saraf, P.}, \bibinfo{author}{Cao, Y.} \&
  \bibinfo{author}{Ramakrishnan, N.}
\newblock \bibinfo{title}{Epidemiological modeling of news and rumors on
  {T}witter}.
\newblock In \emph{\bibinfo{booktitle}{Proceedings of the 7th workshop on
  social network mining and analysis}}, \bibinfo{pages}{1--9}
  (\bibinfo{year}{2013}).

\bibitem{weng2013virality}
\bibinfo{author}{Weng, L.}, \bibinfo{author}{Menczer, F.} \&
  \bibinfo{author}{Ahn, Y.-Y.}
\newblock \bibinfo{journal}{\bibinfo{title}{Virality prediction and community
  structure in social networks}}.
\newblock {\emph{\JournalTitle{Scientific Reports}}}
  \textbf{\bibinfo{volume}{3}}, \bibinfo{pages}{1--6} (\bibinfo{year}{2013}).

\bibitem{bettencourt2006power}
\bibinfo{author}{Bettencourt, L.~M.}, \bibinfo{author}{Cintr{\'o}n-Arias, A.},
  \bibinfo{author}{Kaiser, D.~I.} \& \bibinfo{author}{Castillo-Ch{\'a}vez, C.}
\newblock \bibinfo{journal}{\bibinfo{title}{The power of a good idea:
  Quantitative modeling of the spread of ideas from epidemiological models}}.
\newblock {\emph{\JournalTitle{Physica A}}} \textbf{\bibinfo{volume}{364}},
  \bibinfo{pages}{513--536} (\bibinfo{year}{2006}).

\bibitem{rogers2003diffusion}
\bibinfo{author}{Rogers, E.~M.}
\newblock \emph{\bibinfo{title}{Diffusion of Innovations}}
  (\bibinfo{publisher}{Free Press}, \bibinfo{address}{New York},
  \bibinfo{year}{2003}), \bibinfo{edition}{5th} edn.

\bibitem{dietz1967epidemics}
\bibinfo{author}{Dietz, K.}
\newblock \bibinfo{journal}{\bibinfo{title}{Epidemics and rumours: A survey}}.
\newblock {\emph{\JournalTitle{Journal of the Royal Statistical Society: Series
  A}}} \textbf{\bibinfo{volume}{130}}, \bibinfo{pages}{505--528}
  (\bibinfo{year}{1967}).

\bibitem{van2022misinformation}
\bibinfo{author}{Van Der~Linden, S.}
\newblock \bibinfo{journal}{\bibinfo{title}{Misinformation: susceptibility,
  spread, and interventions to immunize the public}}.
\newblock {\emph{\JournalTitle{Nature Medicine}}}
  \textbf{\bibinfo{volume}{28}}, \bibinfo{pages}{460--467}
  (\bibinfo{year}{2022}).

\bibitem{lovato2024diverse}
\bibinfo{author}{Lovato, J.} \emph{et~al.}
\newblock \bibinfo{journal}{\bibinfo{title}{Diverse misinformation: impacts of
  human biases on detection of deepfakes on networks}}.
\newblock {\emph{\JournalTitle{npj Complexity}}} \textbf{\bibinfo{volume}{1}},
  \bibinfo{pages}{5} (\bibinfo{year}{2024}).

\bibitem{mcadam1993specifying}
\bibinfo{author}{McAdam, D.} \& \bibinfo{author}{Paulsen, R.}
\newblock \bibinfo{journal}{\bibinfo{title}{Specifying the relationship between
  social ties and activism}}.
\newblock {\emph{\JournalTitle{American Journal of Sociology}}}
  \textbf{\bibinfo{volume}{99}}, \bibinfo{pages}{640--667}
  (\bibinfo{year}{1993}).

\bibitem{oliver1998diffusion}
\bibinfo{author}{Oliver, P.~E.} \& \bibinfo{author}{Myers, D.~J.}
\newblock \bibinfo{title}{Diffusion models of cycles of protest as a theory of
  social movements}.
\newblock In \emph{\bibinfo{booktitle}{Congress of the International
  Sociological Association}} (\bibinfo{organization}{Citeseer},
  \bibinfo{year}{1998}).

\bibitem{marsden1998memetics}
\bibinfo{author}{Marsden, P.}
\newblock \bibinfo{journal}{\bibinfo{title}{Memetics and social contagion: Two
  sides of the same coin}}.
\newblock {\emph{\JournalTitle{Journal of Memetics-Evolutionary Models of
  Information Transmission}}} \textbf{\bibinfo{volume}{2}},
  \bibinfo{pages}{171--185} (\bibinfo{year}{1998}).

\bibitem{rivers1937viruses}
\bibinfo{author}{Rivers, T.~M.}
\newblock \bibinfo{journal}{\bibinfo{title}{Viruses and {K}och's postulates}}.
\newblock {\emph{\JournalTitle{Journal of Bacteriology}}}
  \textbf{\bibinfo{volume}{33}}, \bibinfo{pages}{1} (\bibinfo{year}{1937}).

\bibitem{byrd2016adapting}
\bibinfo{author}{Byrd, A.~L.} \& \bibinfo{author}{Segre, J.~A.}
\newblock \bibinfo{journal}{\bibinfo{title}{Adapting {K}och's postulates}}.
\newblock {\emph{\JournalTitle{Science}}} \textbf{\bibinfo{volume}{351}},
  \bibinfo{pages}{224--226} (\bibinfo{year}{2016}).

\bibitem{pirofski2012q}
\bibinfo{author}{Pirofski, L.-A.} \& \bibinfo{author}{Casadevall, A.}
\newblock \bibinfo{journal}{\bibinfo{title}{{Q\&A: What is a pathogen? A
  question that begs the point}}}.
\newblock {\emph{\JournalTitle{BMC Biology}}} \textbf{\bibinfo{volume}{10}},
  \bibinfo{pages}{6} (\bibinfo{year}{2012}).

\bibitem{buffie2013microbiota}
\bibinfo{author}{Buffie, C.~G.} \& \bibinfo{author}{Pamer, E.~G.}
\newblock \bibinfo{journal}{\bibinfo{title}{Microbiota-mediated colonization
  resistance against intestinal pathogens}}.
\newblock {\emph{\JournalTitle{Nature Reviews Immunology}}}
  \textbf{\bibinfo{volume}{13}}, \bibinfo{pages}{790--801}
  (\bibinfo{year}{2013}).

\bibitem{williams2005sickle}
\bibinfo{author}{Williams, T.~N.} \emph{et~al.}
\newblock \bibinfo{journal}{\bibinfo{title}{Sickle cell trait and the risk of
  plasmodium falciparum malaria and other childhood diseases}}.
\newblock {\emph{\JournalTitle{The Journal of Infectious Diseases}}}
  \textbf{\bibinfo{volume}{192}}, \bibinfo{pages}{178--186}
  (\bibinfo{year}{2005}).

\bibitem{brown2012evolution}
\bibinfo{author}{Brown, S.~P.}, \bibinfo{author}{Cornforth, D.~M.} \&
  \bibinfo{author}{Mideo, N.}
\newblock \bibinfo{journal}{\bibinfo{title}{Evolution of virulence in
  opportunistic pathogens: generalism, plasticity, and control}}.
\newblock {\emph{\JournalTitle{Trends in Microbiology}}}
  \textbf{\bibinfo{volume}{20}}, \bibinfo{pages}{336--342}
  (\bibinfo{year}{2012}).

\bibitem{lehmann2018complex}
\bibinfo{author}{Lehmann, S.} \& \bibinfo{author}{Ahn, Y.-Y.}
\newblock \emph{\bibinfo{title}{Complex spreading phenomena in social systems}}
  (\bibinfo{publisher}{Springer}, \bibinfo{year}{2018}).

\bibitem{kata2012anti}
\bibinfo{author}{Kata, A.}
\newblock \bibinfo{journal}{\bibinfo{title}{Anti-vaccine activists, web 2.0,
  and the postmodern paradigm--an overview of tactics and tropes used online by
  the anti-vaccination movement}}.
\newblock {\emph{\JournalTitle{Vaccine}}} \textbf{\bibinfo{volume}{30}},
  \bibinfo{pages}{3778--3789} (\bibinfo{year}{2012}).

\bibitem{klein2024psychology}
\bibinfo{author}{Klein, O.} \& \bibinfo{author}{Yzerbyt, V.}
\newblock \emph{\bibinfo{title}{The psychology of vaccination}}
  (\bibinfo{publisher}{Taylor \& Francis}, \bibinfo{year}{2024}).

\bibitem{chen2020tracking}
\bibinfo{author}{Chen, E.}, \bibinfo{author}{Lerman, K.} \&
  \bibinfo{author}{Ferrara, E.}
\newblock \bibinfo{journal}{\bibinfo{title}{{Tracking social media discourse
  about the COVID-19 pandemic: Development of a public coronavirus Twitter data
  set}}}.
\newblock {\emph{\JournalTitle{JMIR Public Health and Surveillance}}}
  \textbf{\bibinfo{volume}{6}}, \bibinfo{pages}{e19273} (\bibinfo{year}{2020}).

\bibitem{guilbeault2018complex}
\bibinfo{author}{Guilbeault, D.}, \bibinfo{author}{Becker, J.} \&
  \bibinfo{author}{Centola, D.}
\newblock \bibinfo{title}{Complex contagions: A decade in review}.
\newblock In \emph{\bibinfo{booktitle}{Complex Spreading Phenomena in Social
  Systems}}, \bibinfo{pages}{3--25} (\bibinfo{publisher}{Springer},
  \bibinfo{year}{2018}).

\bibitem{hammond2007exploring}
\bibinfo{author}{Hammond, R.~A.} \& \bibinfo{author}{Epstein, J.~M.}
\newblock \bibinfo{journal}{\bibinfo{title}{Exploring price-independent
  mechanisms in the obesity epidemic}}.
\newblock {\emph{\JournalTitle{Center on Social and Economic Dynamics Working
  Paper}}}  (\bibinfo{year}{2007}).

\bibitem{alshamsi2015beyond}
\bibinfo{author}{Alshamsi, A.}, \bibinfo{author}{Pianesi, F.},
  \bibinfo{author}{Lepri, B.}, \bibinfo{author}{Pentland, A.} \&
  \bibinfo{author}{Rahwan, I.}
\newblock \bibinfo{journal}{\bibinfo{title}{Beyond contagion: Reality mining
  reveals complex patterns of social influence}}.
\newblock {\emph{\JournalTitle{PloS One}}} \textbf{\bibinfo{volume}{10}},
  \bibinfo{pages}{e0135740} (\bibinfo{year}{2015}).

\bibitem{galesic2023beyond}
\bibinfo{author}{Galesic, M.} \emph{et~al.}
\newblock \bibinfo{journal}{\bibinfo{title}{Beyond collective intelligence:
  Collective adaptation}}.
\newblock {\emph{\JournalTitle{Journal of the Royal Society Interface}}}
  \textbf{\bibinfo{volume}{20}}, \bibinfo{pages}{20220736}
  (\bibinfo{year}{2023}).

\bibitem{negro2023hepatitis}
\bibinfo{author}{Negro, F.} \& \bibinfo{author}{Lok, A.~S.}
\newblock \bibinfo{journal}{\bibinfo{title}{{Hepatitis D}: a review}}.
\newblock {\emph{\JournalTitle{JAMA}}} \textbf{\bibinfo{volume}{330}},
  \bibinfo{pages}{2376--2387} (\bibinfo{year}{2023}).

\bibitem{rowe2019direct}
\bibinfo{author}{Rowe, H.~M.} \emph{et~al.}
\newblock \bibinfo{journal}{\bibinfo{title}{Direct interactions with influenza
  promote bacterial adherence during respiratory infections}}.
\newblock {\emph{\JournalTitle{Nature Microbiology}}}
  \textbf{\bibinfo{volume}{4}}, \bibinfo{pages}{1328--1336}
  (\bibinfo{year}{2019}).

\bibitem{lian2022bacterial}
\bibinfo{author}{Lian, S.}, \bibinfo{author}{Liu, J.}, \bibinfo{author}{Wu,
  Y.}, \bibinfo{author}{Xia, P.} \& \bibinfo{author}{Zhu, G.}
\newblock \bibinfo{journal}{\bibinfo{title}{Bacterial and viral co-infection in
  the intestine: competition scenario and their effect on host immunity}}.
\newblock {\emph{\JournalTitle{International Journal of Molecular Sciences}}}
  \textbf{\bibinfo{volume}{23}}, \bibinfo{pages}{2311} (\bibinfo{year}{2022}).

\bibitem{seekatz2022role}
\bibinfo{author}{Seekatz, A.~M.}, \bibinfo{author}{Safdar, N.} \&
  \bibinfo{author}{Khanna, S.}
\newblock \bibinfo{journal}{\bibinfo{title}{The role of the gut microbiome in
  colonization resistance and recurrent clostridioides difficile infection}}.
\newblock {\emph{\JournalTitle{Therapeutic Advances in Gastroenterology}}}
  \textbf{\bibinfo{volume}{15}}, \bibinfo{pages}{17562848221134396}
  (\bibinfo{year}{2022}).

\bibitem{martinson2024prevalence}
\bibinfo{author}{Martinson, M.~L.} \& \bibinfo{author}{Lapham, J.}
\newblock \bibinfo{journal}{\bibinfo{title}{Prevalence of immunosuppression
  among {US} adults}}.
\newblock {\emph{\JournalTitle{JAMA}}} \textbf{\bibinfo{volume}{331}},
  \bibinfo{pages}{880--882} (\bibinfo{year}{2024}).

\bibitem{wallace2021prevalence}
\bibinfo{author}{Wallace, B.~I.} \emph{et~al.}
\newblock \bibinfo{journal}{\bibinfo{title}{Prevalence of immunosuppressive
  drug use among commercially insured {US} adults, 2018-2019}}.
\newblock {\emph{\JournalTitle{JAMA Network Open}}}
  \textbf{\bibinfo{volume}{4}}, \bibinfo{pages}{e214920--e214920}
  (\bibinfo{year}{2021}).

\bibitem{morales2023effects}
\bibinfo{author}{Morales, F.}, \bibinfo{author}{Montserrat-De~la Paz, S.},
  \bibinfo{author}{Leon, M.~J.} \& \bibinfo{author}{Rivero-Pino, F.}
\newblock \bibinfo{journal}{\bibinfo{title}{Effects of malnutrition on the
  immune system and infection and the role of nutritional strategies regarding
  improvements in children’s health status: A literature review}}.
\newblock {\emph{\JournalTitle{Nutrients}}} \textbf{\bibinfo{volume}{16}},
  \bibinfo{pages}{1} (\bibinfo{year}{2023}).

\bibitem{bell2018pathogenesis}
\bibinfo{author}{Bell, L.~C.} \& \bibinfo{author}{Noursadeghi, M.}
\newblock \bibinfo{journal}{\bibinfo{title}{{Pathogenesis of HIV-1 and
  Mycobacterium tuberculosis co-infection}}}.
\newblock {\emph{\JournalTitle{Nature Reviews Microbiology}}}
  \textbf{\bibinfo{volume}{16}}, \bibinfo{pages}{80--90}
  (\bibinfo{year}{2018}).

\bibitem{fleming1999epidemiological}
\bibinfo{author}{Fleming, D.~T.} \& \bibinfo{author}{Wasserheit, J.~N.}
\newblock \bibinfo{journal}{\bibinfo{title}{From epidemiological synergy to
  public health policy and practice: the contribution of other sexually
  transmitted diseases to sexual transmission of {HIV} infection.}}
\newblock {\emph{\JournalTitle{Sexually transmitted infections}}}
  \textbf{\bibinfo{volume}{75}}, \bibinfo{pages}{3--17} (\bibinfo{year}{1999}).

\bibitem{do2022infodemics}
\bibinfo{author}{Do~Nascimento, I. J.~B.} \emph{et~al.}
\newblock \bibinfo{journal}{\bibinfo{title}{Infodemics and health
  misinformation: a systematic review of reviews}}.
\newblock {\emph{\JournalTitle{Bulletin of the World Health Organization}}}
  \textbf{\bibinfo{volume}{100}}, \bibinfo{pages}{544} (\bibinfo{year}{2022}).

\bibitem{jin2024social}
\bibinfo{author}{Jin, S.~L.} \emph{et~al.}
\newblock \bibinfo{journal}{\bibinfo{title}{Social histories of public health
  misinformation and infodemics: case studies of four pandemics}}.
\newblock {\emph{\JournalTitle{The Lancet Infectious Diseases}}}
  \textbf{\bibinfo{volume}{24}}, \bibinfo{pages}{e638--e646}
  (\bibinfo{year}{2024}).

\bibitem{dyer2019philippines}
\bibinfo{author}{Dyer, O.}
\newblock \bibinfo{journal}{\bibinfo{title}{Philippines measles outbreak is
  deadliest yet as vaccine scepticism spurs disease comeback}}.
\newblock {\emph{\JournalTitle{BMJ}}} \textbf{\bibinfo{volume}{364}}
  (\bibinfo{year}{2019}).

\bibitem{Normillesafety}
\bibinfo{author}{Normile, D.}
\newblock \bibinfo{journal}{\bibinfo{title}{Safety concerns derail dengue
  vaccination program}}.
\newblock {\emph{\JournalTitle{Science}}} \textbf{\bibinfo{volume}{358}},
  \bibinfo{pages}{1514--1515} (\bibinfo{year}{2017}).

\bibitem{BBC2019}
\bibinfo{author}{{BBC News}}.
\newblock \bibinfo{journal}{\bibinfo{title}{How a wrong injection helped cause
  {Samoa}'s measles epidemic.}}
\newblock {\emph{\JournalTitle{BBC}}} .

\bibitem{champredon2020curbing}
\bibinfo{author}{Champredon, D.}, \bibinfo{author}{Shoukat, A.},
  \bibinfo{author}{Singer, B.~H.}, \bibinfo{author}{Galvani, A.~P.} \&
  \bibinfo{author}{Moghadas, S.~M.}
\newblock \bibinfo{journal}{\bibinfo{title}{Curbing the 2019 {Samoa} measles
  outbreak}}.
\newblock {\emph{\JournalTitle{The Lancet Infectious Diseases}}}
  \textbf{\bibinfo{volume}{20}}, \bibinfo{pages}{287--288}
  (\bibinfo{year}{2020}).

\bibitem{grant2013ingesting}
\bibinfo{author}{Grant-Alfieri, A.}, \bibinfo{author}{Schaechter, J.} \&
  \bibinfo{author}{Lipshultz, S.~E.}
\newblock \bibinfo{journal}{\bibinfo{title}{Ingesting and aspirating dry
  cinnamon by children and adolescents: the “cinnamon challenge”}}.
\newblock {\emph{\JournalTitle{Pediatrics}}} \textbf{\bibinfo{volume}{131}},
  \bibinfo{pages}{833--835} (\bibinfo{year}{2013}).

\bibitem{breakey2015salt}
\bibinfo{author}{Breakey, W.}, \bibinfo{author}{Crowley, T.~P.} \&
  \bibinfo{author}{Alrawi, M.}
\newblock \bibinfo{journal}{\bibinfo{title}{Salt and ice, a challenge not to be
  taken lightly}}.
\newblock {\emph{\JournalTitle{Journal of Burn Care \& Research}}}
  \textbf{\bibinfo{volume}{36}}, \bibinfo{pages}{e230--e230}
  (\bibinfo{year}{2015}).

\bibitem{FDA2020}
\bibinfo{author}{{Drug Safety Communication}}.
\newblock \bibinfo{journal}{\bibinfo{title}{Fda warns about serious problems
  with high doses of the allergy medicine diphenhydramine (benadryl)}}.
\newblock {\emph{\JournalTitle{FDA}}} .

\bibitem{merchant2021public}
\bibinfo{author}{Merchant, R.~M.}, \bibinfo{author}{South, E.~C.} \&
  \bibinfo{author}{Lurie, N.}
\newblock \bibinfo{journal}{\bibinfo{title}{Public health messaging in an era
  of social media}}.
\newblock {\emph{\JournalTitle{JAMA}}} \textbf{\bibinfo{volume}{325}},
  \bibinfo{pages}{223--224} (\bibinfo{year}{2021}).

\bibitem{hyland2021toward}
\bibinfo{author}{Hyland-Wood, B.}, \bibinfo{author}{Gardner, J.},
  \bibinfo{author}{Leask, J.} \& \bibinfo{author}{Ecker, U.~K.}
\newblock \bibinfo{journal}{\bibinfo{title}{{Toward effective government
  communication strategies in the era of COVID-19}}}.
\newblock {\emph{\JournalTitle{Humanities and Social Sciences Communications}}}
  \textbf{\bibinfo{volume}{8}} (\bibinfo{year}{2021}).

\bibitem{krapivsky2024epidemic}
\bibinfo{author}{Krapivsky, P.} \& \bibinfo{author}{Redner, S.}
\newblock \bibinfo{journal}{\bibinfo{title}{Epidemic forecast follies}}.
\newblock {\emph{\JournalTitle{npj Complexity}}} \textbf{\bibinfo{volume}{1}},
  \bibinfo{pages}{7} (\bibinfo{year}{2024}).

\bibitem{pathak2010generalized}
\bibinfo{author}{Pathak, N.}, \bibinfo{author}{Banerjee, A.} \&
  \bibinfo{author}{Srivastava, J.}
\newblock \bibinfo{title}{A generalized linear threshold model for multiple
  cascades}.
\newblock In \emph{\bibinfo{booktitle}{2010 IEEE International Conference on
  Data Mining}}, \bibinfo{pages}{965--970} (\bibinfo{organization}{IEEE},
  \bibinfo{year}{2010}).

\bibitem{mehta2020modelling}
\bibinfo{author}{Mehta, R.~S.} \& \bibinfo{author}{Rosenberg, N.~A.}
\newblock \bibinfo{journal}{\bibinfo{title}{Modelling anti-vaccine sentiment as
  a cultural pathogen}}.
\newblock {\emph{\JournalTitle{Evolutionary Human Sciences}}}
  \textbf{\bibinfo{volume}{2}}, \bibinfo{pages}{e21} (\bibinfo{year}{2020}).

\bibitem{sznajd2005left}
\bibinfo{author}{Sznajd-Weron, K.} \& \bibinfo{author}{Sznajd, J.}
\newblock \bibinfo{journal}{\bibinfo{title}{Who is left, who is right?}}
\newblock {\emph{\JournalTitle{Physica A}}} \textbf{\bibinfo{volume}{351}},
  \bibinfo{pages}{593--604} (\bibinfo{year}{2005}).

\bibitem{fu2017dueling}
\bibinfo{author}{Fu, F.}, \bibinfo{author}{Christakis, N.~A.} \&
  \bibinfo{author}{Fowler, J.~H.}
\newblock \bibinfo{journal}{\bibinfo{title}{Dueling biological and social
  contagions}}.
\newblock {\emph{\JournalTitle{Scientific Reports}}}
  \textbf{\bibinfo{volume}{7}}, \bibinfo{pages}{43634} (\bibinfo{year}{2017}).

\bibitem{tambuscio2015fact}
\bibinfo{author}{Tambuscio, M.}, \bibinfo{author}{Ruffo, G.},
  \bibinfo{author}{Flammini, A.} \& \bibinfo{author}{Menczer, F.}
\newblock \bibinfo{title}{Fact-checking effect on viral hoaxes: A model of
  misinformation spread in social networks}.
\newblock In \emph{\bibinfo{booktitle}{Proceedings of the 24th international
  conference on World Wide Web}}, \bibinfo{pages}{977--982}
  (\bibinfo{year}{2015}).

\bibitem{hebert2020spread}
\bibinfo{author}{H{\'e}bert-Dufresne, L.}, \bibinfo{author}{Mistry, D.} \&
  \bibinfo{author}{Althouse, B.~M.}
\newblock \bibinfo{journal}{\bibinfo{title}{Spread of infectious disease and
  social awareness as parasitic contagions on clustered networks}}.
\newblock {\emph{\JournalTitle{Physical Review Research}}}
  \textbf{\bibinfo{volume}{2}}, \bibinfo{pages}{033306} (\bibinfo{year}{2020}).

\bibitem{briand2021infodemics}
\bibinfo{author}{Briand, S.~C.} \emph{et~al.}
\newblock \bibinfo{journal}{\bibinfo{title}{Infodemics: A new challenge for
  public health}}.
\newblock {\emph{\JournalTitle{Cell}}} \textbf{\bibinfo{volume}{184}},
  \bibinfo{pages}{6010--6014} (\bibinfo{year}{2021}).

\bibitem{bedson2021review}
\bibinfo{author}{Bedson, J.} \emph{et~al.}
\newblock \bibinfo{journal}{\bibinfo{title}{A review and agenda for integrated
  disease models including social and behavioural factors}}.
\newblock {\emph{\JournalTitle{Nature Human Behaviour}}}
  \textbf{\bibinfo{volume}{5}}, \bibinfo{pages}{834--846}
  (\bibinfo{year}{2021}).

\bibitem{anderson1972more}
\bibinfo{author}{Anderson, P.~W.}
\newblock \bibinfo{journal}{\bibinfo{title}{More is different: Broken symmetry
  and the nature of the hierarchical structure of science.}}
\newblock {\emph{\JournalTitle{Science}}} \textbf{\bibinfo{volume}{177}},
  \bibinfo{pages}{393--396} (\bibinfo{year}{1972}).

\bibitem{hebert2024path}
\bibinfo{author}{H{\'e}bert-Dufresne, L.}, \bibinfo{author}{Allard, A.},
  \bibinfo{author}{Garland, J.}, \bibinfo{author}{Hobson, E.~A.} \&
  \bibinfo{author}{Zaman, L.}
\newblock \bibinfo{journal}{\bibinfo{title}{The path of complexity}}.
\newblock {\emph{\JournalTitle{npj Complexity}}} \textbf{\bibinfo{volume}{1}},
  \bibinfo{pages}{4} (\bibinfo{year}{2024}).

\bibitem{bernoulli1760essai}
\bibinfo{author}{Bernoulli, D.}
\newblock \bibinfo{journal}{\bibinfo{title}{Essai d'une nouvelle analyse de la
  mortalit{\'e} caus{\'e}e par la petite v{\'e}role, et des avantages de
  l'inoculation pour la pr{\'e}venir}}.
\newblock  (\bibinfo{year}{1760}).

\bibitem{kermack1927a}
\bibinfo{author}{Kermack, W.~O.} \& \bibinfo{author}{Mc{K}endrick, A.~G.}
\newblock \bibinfo{journal}{\bibinfo{title}{A contribution to the mathematical
  theory of epidemics}}.
\newblock {\emph{\JournalTitle{Proc. R. Soc. Lond. A}}}
  \textbf{\bibinfo{volume}{115}}, \bibinfo{pages}{700--721}
  (\bibinfo{year}{1927}).

\bibitem{may1995coinfection}
\bibinfo{author}{May, R.~M.} \& \bibinfo{author}{Nowak, M.~A.}
\newblock \bibinfo{journal}{\bibinfo{title}{Coinfection and the evolution of
  parasite virulence}}.
\newblock {\emph{\JournalTitle{Proceedings of the Royal Society of London.
  Series B}}} \textbf{\bibinfo{volume}{261}}, \bibinfo{pages}{209--215}
  (\bibinfo{year}{1995}).

\bibitem{andreasen1997dynamics}
\bibinfo{author}{Andreasen, V.}, \bibinfo{author}{Lin, J.} \&
  \bibinfo{author}{Levin, S.~A.}
\newblock \bibinfo{journal}{\bibinfo{title}{The dynamics of cocirculating
  influenza strains conferring partial cross-immunity}}.
\newblock {\emph{\JournalTitle{Journal of Mathematical Biology}}}
  \textbf{\bibinfo{volume}{35}}, \bibinfo{pages}{825--842}
  (\bibinfo{year}{1997}).

\bibitem{epstein2008coupled}
\bibinfo{author}{Epstein, J.~M.}, \bibinfo{author}{Parker, J.},
  \bibinfo{author}{Cummings, D.} \& \bibinfo{author}{Hammond, R.~A.}
\newblock \bibinfo{journal}{\bibinfo{title}{Coupled contagion dynamics of fear
  and disease: mathematical and computational explorations}}.
\newblock {\emph{\JournalTitle{PloS One}}} \textbf{\bibinfo{volume}{3}},
  \bibinfo{pages}{e3955} (\bibinfo{year}{2008}).

\bibitem{newman2013interacting}
\bibinfo{author}{Newman, M.~E.} \& \bibinfo{author}{Ferrario, C.~R.}
\newblock \bibinfo{journal}{\bibinfo{title}{Interacting epidemics and
  coinfection on contact networks}}.
\newblock {\emph{\JournalTitle{PloS One}}} \textbf{\bibinfo{volume}{8}},
  \bibinfo{pages}{e71321} (\bibinfo{year}{2013}).

\bibitem{gog2002dynamics}
\bibinfo{author}{Gog, J.~R.} \& \bibinfo{author}{Grenfell, B.~T.}
\newblock \bibinfo{journal}{\bibinfo{title}{Dynamics and selection of
  many-strain pathogens}}.
\newblock {\emph{\JournalTitle{Proceedings of the National Academy of
  Sciences}}} \textbf{\bibinfo{volume}{99}}, \bibinfo{pages}{17209--17214}
  (\bibinfo{year}{2002}).

\bibitem{karrer2011competing}
\bibinfo{author}{Karrer, B.} \& \bibinfo{author}{Newman, M.~E.}
\newblock \bibinfo{journal}{\bibinfo{title}{Competing epidemics on complex
  networks}}.
\newblock {\emph{\JournalTitle{Physical Review E}}}
  \textbf{\bibinfo{volume}{84}}, \bibinfo{pages}{036106}
  (\bibinfo{year}{2011}).

\bibitem{bansal2012impact}
\bibinfo{author}{Bansal, S.} \& \bibinfo{author}{Meyers, L.~A.}
\newblock \bibinfo{journal}{\bibinfo{title}{The impact of past epidemics on
  future disease dynamics}}.
\newblock {\emph{\JournalTitle{Journal of Theoretical Biology}}}
  \textbf{\bibinfo{volume}{309}}, \bibinfo{pages}{176--184}
  (\bibinfo{year}{2012}).

\bibitem{miller2013cocirculation}
\bibinfo{author}{Miller, J.~C.}
\newblock \bibinfo{journal}{\bibinfo{title}{Cocirculation of infectious
  diseases on networks}}.
\newblock {\emph{\JournalTitle{Physical Review E}}}
  \textbf{\bibinfo{volume}{87}}, \bibinfo{pages}{060801}
  (\bibinfo{year}{2013}).

\bibitem{poletto2013host}
\bibinfo{author}{Poletto, C.}, \bibinfo{author}{Meloni, S.},
  \bibinfo{author}{Colizza, V.}, \bibinfo{author}{Moreno, Y.} \&
  \bibinfo{author}{Vespignani, A.}
\newblock \bibinfo{journal}{\bibinfo{title}{Host mobility drives pathogen
  competition in spatially structured populations}}.
\newblock {\emph{\JournalTitle{PLoS Computational Biology}}}
  \textbf{\bibinfo{volume}{9}}, \bibinfo{pages}{e1003169}
  (\bibinfo{year}{2013}).

\bibitem{gleeson2014competition}
\bibinfo{author}{Gleeson, J.~P.}, \bibinfo{author}{Ward, J.~A.},
  \bibinfo{author}{{O'Sullivan}, K.~P.} \& \bibinfo{author}{Lee, W.~T.}
\newblock \bibinfo{journal}{\bibinfo{title}{Competition-induced criticality in
  a model of meme popularity}}.
\newblock {\emph{\JournalTitle{Physical Review Letters}}}
  \textbf{\bibinfo{volume}{112}}, \bibinfo{pages}{048701}
  (\bibinfo{year}{2014}).

\bibitem{williams2022immunity}
\bibinfo{author}{Williams, B.~J.}, \bibinfo{author}{Ogbunugafor, C.~B.},
  \bibinfo{author}{Althouse, B.~M.} \& \bibinfo{author}{H{\'e}bert-Dufresne,
  L.}
\newblock \bibinfo{journal}{\bibinfo{title}{Immunity-induced criticality of the
  genotype network of influenza {A (H3N2)} hemagglutinin}}.
\newblock {\emph{\JournalTitle{PNAS Nexus}}} \textbf{\bibinfo{volume}{1}},
  \bibinfo{pages}{pgac143} (\bibinfo{year}{2022}).

\bibitem{newman2005threshold}
\bibinfo{author}{Newman, M.~E.}
\newblock \bibinfo{journal}{\bibinfo{title}{Threshold effects for two pathogens
  spreading on a network}}.
\newblock {\emph{\JournalTitle{Physical Review Letters}}}
  \textbf{\bibinfo{volume}{95}}, \bibinfo{pages}{108701}
  (\bibinfo{year}{2005}).

\bibitem{funk2009spread}
\bibinfo{author}{Funk, S.}, \bibinfo{author}{Gilad, E.},
  \bibinfo{author}{Watkins, C.} \& \bibinfo{author}{Jansen, V.~A.}
\newblock \bibinfo{journal}{\bibinfo{title}{The spread of awareness and its
  impact on epidemic outbreaks}}.
\newblock {\emph{\JournalTitle{Proceedings of the National Academy of
  Sciences}}} \textbf{\bibinfo{volume}{106}}, \bibinfo{pages}{6872--6877}
  (\bibinfo{year}{2009}).

\bibitem{funk2010interacting}
\bibinfo{author}{Funk, S.} \& \bibinfo{author}{Jansen, V.~A.}
\newblock \bibinfo{journal}{\bibinfo{title}{Interacting epidemics on overlay
  networks}}.
\newblock {\emph{\JournalTitle{Physical Review E}}}
  \textbf{\bibinfo{volume}{81}}, \bibinfo{pages}{036118}
  (\bibinfo{year}{2010}).

\bibitem{marceau2011modeling}
\bibinfo{author}{Marceau, V.}, \bibinfo{author}{No{\"e}l, P.-A.},
  \bibinfo{author}{H{\'e}bert-Dufresne, L.}, \bibinfo{author}{Allard, A.} \&
  \bibinfo{author}{Dub{\'e}, L.~J.}
\newblock \bibinfo{journal}{\bibinfo{title}{Modeling the dynamical interaction
  between epidemics on overlay networks}}.
\newblock {\emph{\JournalTitle{Physical Review E}}}
  \textbf{\bibinfo{volume}{84}}, \bibinfo{pages}{026105}
  (\bibinfo{year}{2011}).

\bibitem{wang2019coevolution}
\bibinfo{author}{Wang, W.}, \bibinfo{author}{Liu, Q.-H.},
  \bibinfo{author}{Liang, J.}, \bibinfo{author}{Hu, Y.} \&
  \bibinfo{author}{Zhou, T.}
\newblock \bibinfo{journal}{\bibinfo{title}{Coevolution spreading in complex
  networks}}.
\newblock {\emph{\JournalTitle{Physics Reports}}}
  \textbf{\bibinfo{volume}{820}}, \bibinfo{pages}{1--51}
  (\bibinfo{year}{2019}).

\bibitem{vasco2007tracking}
\bibinfo{author}{Vasco, D.~A.}, \bibinfo{author}{Wearing, H.~J.} \&
  \bibinfo{author}{Rohani, P.}
\newblock \bibinfo{journal}{\bibinfo{title}{Tracking the dynamics of pathogen
  interactions: modeling ecological and immune-mediated processes in a
  two-pathogen single-host system}}.
\newblock {\emph{\JournalTitle{Journal of Theoretical Biology}}}
  \textbf{\bibinfo{volume}{245}}, \bibinfo{pages}{9--25}
  (\bibinfo{year}{2007}).

\bibitem{abu2008interactions}
\bibinfo{author}{Abu-Raddad, L.~J.}, \bibinfo{author}{Van~der Ventel, B.} \&
  \bibinfo{author}{Ferguson, N.~M.}
\newblock \bibinfo{journal}{\bibinfo{title}{Interactions of multiple strain
  pathogen diseases in the presence of coinfection, cross immunity, and
  arbitrary strain diversity}}.
\newblock {\emph{\JournalTitle{Physical Review Letters}}}
  \textbf{\bibinfo{volume}{100}}, \bibinfo{pages}{168102}
  (\bibinfo{year}{2008}).

\bibitem{hebert2013pathogen}
\bibinfo{author}{H{\'e}bert-Dufresne, L.}, \bibinfo{author}{Patterson-Lomba,
  O.}, \bibinfo{author}{Goerg, G.~M.} \& \bibinfo{author}{Althouse, B.~M.}
\newblock \bibinfo{journal}{\bibinfo{title}{Pathogen mutation modeled by
  competition between site and bond percolation}}.
\newblock {\emph{\JournalTitle{Physical Review Letters}}}
  \textbf{\bibinfo{volume}{110}}, \bibinfo{pages}{108103}
  (\bibinfo{year}{2013}).

\bibitem{campbell2013complex}
\bibinfo{author}{Campbell, E.} \& \bibinfo{author}{Salath{\'e}, M.}
\newblock \bibinfo{journal}{\bibinfo{title}{Complex social contagion makes
  networks more vulnerable to disease outbreaks}}.
\newblock {\emph{\JournalTitle{Scientific Reports}}}
  \textbf{\bibinfo{volume}{3}}, \bibinfo{pages}{1905} (\bibinfo{year}{2013}).

\bibitem{de2016physics}
\bibinfo{author}{De~Domenico, M.}, \bibinfo{author}{Granell, C.},
  \bibinfo{author}{Porter, M.~A.} \& \bibinfo{author}{Arenas, A.}
\newblock \bibinfo{journal}{\bibinfo{title}{The physics of spreading processes
  in multilayer networks}}.
\newblock {\emph{\JournalTitle{Nature Physics}}} \textbf{\bibinfo{volume}{12}},
  \bibinfo{pages}{901} (\bibinfo{year}{2016}).

\bibitem{liu2018interactive}
\bibinfo{author}{Liu, Q.-H.}, \bibinfo{author}{Zhong, L.-F.},
  \bibinfo{author}{Wang, W.}, \bibinfo{author}{Zhou, T.} \&
  \bibinfo{author}{Eugene~Stanley, H.}
\newblock \bibinfo{journal}{\bibinfo{title}{Interactive social contagions and
  co-infections on complex networks}}.
\newblock {\emph{\JournalTitle{Chaos}}} \textbf{\bibinfo{volume}{28}}
  (\bibinfo{year}{2018}).

\bibitem{chen2013outbreaks}
\bibinfo{author}{Chen, L.}, \bibinfo{author}{Ghanbarnejad, F.},
  \bibinfo{author}{Cai, W.} \& \bibinfo{author}{Grassberger, P.}
\newblock \bibinfo{journal}{\bibinfo{title}{Outbreaks of coinfections: The
  critical role of cooperativity}}.
\newblock {\emph{\JournalTitle{Europhysics Letters}}}
  \textbf{\bibinfo{volume}{104}}, \bibinfo{pages}{50001}
  (\bibinfo{year}{2013}).

\bibitem{hebert2015complex}
\bibinfo{author}{H{\'e}bert-Dufresne, L.} \& \bibinfo{author}{Althouse, B.~M.}
\newblock \bibinfo{journal}{\bibinfo{title}{Complex dynamics of synergistic
  coinfections on realistically clustered networks}}.
\newblock {\emph{\JournalTitle{Proceedings of the National Academy of
  Sciences}}} \textbf{\bibinfo{volume}{112}}, \bibinfo{pages}{10551--10556}
  (\bibinfo{year}{2015}).

\bibitem{cai2015avalanche}
\bibinfo{author}{Cai, W.}, \bibinfo{author}{Chen, L.},
  \bibinfo{author}{Ghanbarnejad, F.} \& \bibinfo{author}{Grassberger, P.}
\newblock \bibinfo{journal}{\bibinfo{title}{Avalanche outbreaks emerging in
  cooperative contagions}}.
\newblock {\emph{\JournalTitle{Nature Physics}}} \textbf{\bibinfo{volume}{11}},
  \bibinfo{pages}{936} (\bibinfo{year}{2015}).

\bibitem{chen2017fundamental}
\bibinfo{author}{Chen, L.}, \bibinfo{author}{Ghanbarnejad, F.} \&
  \bibinfo{author}{Brockmann, D.}
\newblock \bibinfo{journal}{\bibinfo{title}{Fundamental properties of
  cooperative contagion processes}}.
\newblock {\emph{\JournalTitle{New Journal of Physics}}}
  \textbf{\bibinfo{volume}{19}}, \bibinfo{pages}{103041}
  (\bibinfo{year}{2017}).

\bibitem{soriano2019markovian}
\bibinfo{author}{Soriano-Pa{\~n}os, D.}, \bibinfo{author}{Ghanbarnejad, F.},
  \bibinfo{author}{Meloni, S.} \& \bibinfo{author}{G{\'o}mez-Garde{\~n}es, J.}
\newblock \bibinfo{journal}{\bibinfo{title}{Markovian approach to tackle the
  interaction of simultaneous diseases}}.
\newblock {\emph{\JournalTitle{Physical Review E}}}
  \textbf{\bibinfo{volume}{100}}, \bibinfo{pages}{062308}
  (\bibinfo{year}{2019}).

\bibitem{grassberger2016phase}
\bibinfo{author}{Grassberger, P.}, \bibinfo{author}{Chen, L.},
  \bibinfo{author}{Ghanbarnejad, F.} \& \bibinfo{author}{Cai, W.}
\newblock \bibinfo{journal}{\bibinfo{title}{Phase transitions in cooperative
  coinfections: Simulation results for networks and lattices}}.
\newblock {\emph{\JournalTitle{Physical Review E}}}
  \textbf{\bibinfo{volume}{93}}, \bibinfo{pages}{042316}
  (\bibinfo{year}{2016}).

\bibitem{hebert2020macroscopic}
\bibinfo{author}{H{\'e}bert-Dufresne, L.}, \bibinfo{author}{Scarpino, S.~V.} \&
  \bibinfo{author}{Young, J.-G.}
\newblock \bibinfo{journal}{\bibinfo{title}{Macroscopic patterns of interacting
  contagions are indistinguishable from social reinforcement}}.
\newblock {\emph{\JournalTitle{Nature Physics}}} \textbf{\bibinfo{volume}{16}},
  \bibinfo{pages}{426--431} (\bibinfo{year}{2020}).

\bibitem{lamata2024pathways}
\bibinfo{author}{Lamata-Ot{\'\i}n, S.},
  \bibinfo{author}{G{\'o}mez-Garde{\~n}es, J.} \&
  \bibinfo{author}{Soriano-Pa{\~n}os, D.}
\newblock \bibinfo{journal}{\bibinfo{title}{Pathways to discontinuous
  transitions in interacting contagion dynamics}}.
\newblock {\emph{\JournalTitle{Journal of Physics: Complexity}}}
  \textbf{\bibinfo{volume}{5}}, \bibinfo{pages}{015015} (\bibinfo{year}{2024}).

\bibitem{centola2010spread}
\bibinfo{author}{Centola, D.}
\newblock \bibinfo{journal}{\bibinfo{title}{The spread of behavior in an online
  social network experiment}}.
\newblock {\emph{\JournalTitle{Science}}} \textbf{\bibinfo{volume}{329}},
  \bibinfo{pages}{1194--1197} (\bibinfo{year}{2010}).

\bibitem{de2018fundamentals}
\bibinfo{author}{de~Arruda, G.~F.}, \bibinfo{author}{Rodrigues, F.~A.} \&
  \bibinfo{author}{Moreno, Y.}
\newblock \bibinfo{journal}{\bibinfo{title}{Fundamentals of spreading processes
  in single and multilayer complex networks}}.
\newblock {\emph{\JournalTitle{Physics Reports}}}
  \textbf{\bibinfo{volume}{756}}, \bibinfo{pages}{1--59}
  (\bibinfo{year}{2018}).

\bibitem{wang2024epidemic}
\bibinfo{author}{Wang, W.} \emph{et~al.}
\newblock \bibinfo{journal}{\bibinfo{title}{Epidemic spreading on higher-order
  networks}}.
\newblock {\emph{\JournalTitle{Physics Reports}}}
  \textbf{\bibinfo{volume}{1056}}, \bibinfo{pages}{1--70}
  (\bibinfo{year}{2024}).

\bibitem{brodka2020interacting}
\bibinfo{author}{Br{\'o}dka, P.}, \bibinfo{author}{Musial, K.} \&
  \bibinfo{author}{Jankowski, J.}
\newblock \bibinfo{journal}{\bibinfo{title}{Interacting spreading processes in
  multilayer networks: A systematic review}}.
\newblock {\emph{\JournalTitle{IEEE Access}}} \textbf{\bibinfo{volume}{8}},
  \bibinfo{pages}{10316--10341} (\bibinfo{year}{2020}).

\bibitem{huang2021modeling}
\bibinfo{author}{Huang, H.}, \bibinfo{author}{Chen, Y.} \& \bibinfo{author}{Ma,
  Y.}
\newblock \bibinfo{journal}{\bibinfo{title}{Modeling the competitive diffusions
  of rumor and knowledge and the impacts on epidemic spreading}}.
\newblock {\emph{\JournalTitle{Applied Mathematics and Computation}}}
  \textbf{\bibinfo{volume}{388}}, \bibinfo{pages}{125536}
  (\bibinfo{year}{2021}).

\bibitem{li2022competing}
\bibinfo{author}{Li, W.}, \bibinfo{author}{Xue, X.}, \bibinfo{author}{Pan, L.},
  \bibinfo{author}{Lin, T.} \& \bibinfo{author}{Wang, W.}
\newblock \bibinfo{journal}{\bibinfo{title}{Competing spreading dynamics in
  simplicial complex}}.
\newblock {\emph{\JournalTitle{Applied Mathematics and Computation}}}
  \textbf{\bibinfo{volume}{412}}, \bibinfo{pages}{126595}
  (\bibinfo{year}{2022}).

\bibitem{nie2022markovian}
\bibinfo{author}{Nie, Y.}, \bibinfo{author}{Li, W.}, \bibinfo{author}{Pan, L.},
  \bibinfo{author}{Lin, T.} \& \bibinfo{author}{Wang, W.}
\newblock \bibinfo{journal}{\bibinfo{title}{Markovian approach to tackle
  competing pathogens in simplicial complex}}.
\newblock {\emph{\JournalTitle{Applied Mathematics and Computation}}}
  \textbf{\bibinfo{volume}{417}}, \bibinfo{pages}{126773}
  (\bibinfo{year}{2022}).

\bibitem{li2023coevolution}
\bibinfo{author}{Li, W.} \emph{et~al.}
\newblock \bibinfo{journal}{\bibinfo{title}{Coevolution of epidemic and
  infodemic on higher-order networks}}.
\newblock {\emph{\JournalTitle{Chaos, Solitons \& Fractals}}}
  \textbf{\bibinfo{volume}{168}}, \bibinfo{pages}{113102}
  (\bibinfo{year}{2023}).

\bibitem{fan2022epidemics}
\bibinfo{author}{Fan, J.}, \bibinfo{author}{Yin, Q.}, \bibinfo{author}{Xia, C.}
  \& \bibinfo{author}{Perc, M.}
\newblock \bibinfo{journal}{\bibinfo{title}{Epidemics on multilayer simplicial
  complexes}}.
\newblock {\emph{\JournalTitle{Proceedings of the Royal Society A}}}
  \textbf{\bibinfo{volume}{478}}, \bibinfo{pages}{20220059}
  (\bibinfo{year}{2022}).

\bibitem{you2023impact}
\bibinfo{author}{You, X.}, \bibinfo{author}{Zhang, M.}, \bibinfo{author}{Ma,
  Y.}, \bibinfo{author}{Tan, J.} \& \bibinfo{author}{Liu, Z.}
\newblock \bibinfo{journal}{\bibinfo{title}{Impact of higher-order interactions
  and individual emotional heterogeneity on information-disease coupled
  dynamics in multiplex networks}}.
\newblock {\emph{\JournalTitle{Chaos, Solitons \& Fractals}}}
  \textbf{\bibinfo{volume}{177}}, \bibinfo{pages}{114186}
  (\bibinfo{year}{2023}).

\bibitem{liu2023epidemic}
\bibinfo{author}{Liu, L.}, \bibinfo{author}{Feng, M.}, \bibinfo{author}{Xia,
  C.}, \bibinfo{author}{Zhao, D.} \& \bibinfo{author}{Perc, M.}
\newblock \bibinfo{journal}{\bibinfo{title}{Epidemic trajectories and awareness
  diffusion among unequals in simplicial complexes}}.
\newblock {\emph{\JournalTitle{Chaos, Solitons \& Fractals}}}
  \textbf{\bibinfo{volume}{173}}, \bibinfo{pages}{113657}
  (\bibinfo{year}{2023}).

\bibitem{hong2023coupled}
\bibinfo{author}{Hong, Z.}, \bibinfo{author}{Zhou, H.}, \bibinfo{author}{Wang,
  Z.}, \bibinfo{author}{Yin, Q.} \& \bibinfo{author}{Liu, J.}
\newblock \bibinfo{journal}{\bibinfo{title}{Coupled propagation dynamics of
  information and infectious disease on two-layer complex networks with
  simplices}}.
\newblock {\emph{\JournalTitle{Mathematics}}} \textbf{\bibinfo{volume}{11}},
  \bibinfo{pages}{4904} (\bibinfo{year}{2023}).

\bibitem{zhu2023epidemic}
\bibinfo{author}{Zhu, Y.}, \bibinfo{author}{Li, C.} \& \bibinfo{author}{Li, X.}
\newblock \bibinfo{journal}{\bibinfo{title}{Epidemic spreading on coupling
  network with higher-order information layer}}.
\newblock {\emph{\JournalTitle{New Journal of Physics}}}
  \textbf{\bibinfo{volume}{25}}, \bibinfo{pages}{113043}
  (\bibinfo{year}{2023}).

\bibitem{min2014diversity}
\bibinfo{author}{Min, Y.} \emph{et~al.}
\newblock \bibinfo{journal}{\bibinfo{title}{Diversity of multilayer networks
  and its impact on collaborating epidemics}}.
\newblock {\emph{\JournalTitle{Physical Review E}}}
  \textbf{\bibinfo{volume}{90}}, \bibinfo{pages}{062803}
  (\bibinfo{year}{2014}).

\bibitem{nguyen2024upper}
\bibinfo{author}{Nguyen, M.~M.}
\newblock \bibinfo{journal}{\bibinfo{title}{Upper bounds on overshoot in sir
  models with nonlinear incidence}}.
\newblock {\emph{\JournalTitle{npj Complexity}}} \textbf{\bibinfo{volume}{1}},
  \bibinfo{pages}{11} (\bibinfo{year}{2024}).

\bibitem{katz1955a}
\bibinfo{author}{Katz, E.} \& \bibinfo{author}{Lazarsfeld, P.~F.}
\newblock \emph{\bibinfo{title}{Personal Influence}} (\bibinfo{publisher}{The
  Free Press}, \bibinfo{address}{New York}, \bibinfo{year}{1955}).

\bibitem{weimann1994a}
\bibinfo{author}{Weimann, G.}
\newblock \emph{\bibinfo{title}{The Influentials: People Who Influence People}}
  (\bibinfo{publisher}{State University of New York Press},
  \bibinfo{address}{Albany, NY}, \bibinfo{year}{1994}).

\bibitem{gladwell2000a}
\bibinfo{author}{Gladwell, M.}
\newblock \emph{\bibinfo{title}{The Tipping Point}}
  (\bibinfo{publisher}{Little, Brown and Company}, \bibinfo{address}{New York},
  \bibinfo{year}{2000}).

\bibitem{watts2007a}
\bibinfo{author}{Watts, D.~J.} \& \bibinfo{author}{Dodds, P.~S.}
\newblock \bibinfo{journal}{\bibinfo{title}{Influentials, networks, and public
  opinion formation}}.
\newblock {\emph{\JournalTitle{Journal of Consumer Research}}}
  \textbf{\bibinfo{volume}{34}}, \bibinfo{pages}{441--458}
  (\bibinfo{year}{2007}).

\bibitem{goffman1964a}
\bibinfo{author}{Goffman, W.} \& \bibinfo{author}{Newill, V.~A.}
\newblock \bibinfo{journal}{\bibinfo{title}{Generalization of epidemic theory:
  {A}n application to the transmission of ideas}}.
\newblock {\emph{\JournalTitle{Nature}}} \textbf{\bibinfo{volume}{204}},
  \bibinfo{pages}{225--228} (\bibinfo{year}{1964}).

\bibitem{cavalli1982theory}
\bibinfo{author}{Cavalli-Sforza, L.~L.}, \bibinfo{author}{Feldman, M.~W.},
  \bibinfo{author}{Chen, K.-H.} \& \bibinfo{author}{Dornbusch, S.~M.}
\newblock \bibinfo{journal}{\bibinfo{title}{Theory and observation in cultural
  transmission}}.
\newblock {\emph{\JournalTitle{Science}}} \textbf{\bibinfo{volume}{218}},
  \bibinfo{pages}{19--27} (\bibinfo{year}{1982}).

\bibitem{ananthasubramaniam2024networks}
\bibinfo{author}{Ananthasubramaniam, A.}, \bibinfo{author}{Jurgens, D.} \&
  \bibinfo{author}{Romero, D.~M.}
\newblock \bibinfo{journal}{\bibinfo{title}{Networks and identity drive the
  spatial diffusion of linguistic innovation in urban and rural areas}}.
\newblock {\emph{\JournalTitle{npj Complexity}}} \textbf{\bibinfo{volume}{1}},
  \bibinfo{pages}{14} (\bibinfo{year}{2024}).

\bibitem{lerman2024affective}
\bibinfo{author}{Lerman, K.}, \bibinfo{author}{Feldman, D.},
  \bibinfo{author}{He, Z.} \& \bibinfo{author}{Rao, A.}
\newblock \bibinfo{journal}{\bibinfo{title}{Affective polarization and dynamics
  of information spread in online networks}}.
\newblock {\emph{\JournalTitle{npj Complexity}}} \textbf{\bibinfo{volume}{1}},
  \bibinfo{pages}{8} (\bibinfo{year}{2024}).

\bibitem{schelling1971a}
\bibinfo{author}{Schelling, T.~C.}
\newblock \bibinfo{journal}{\bibinfo{title}{Dynamic models of segregation}}.
\newblock {\emph{\JournalTitle{J. Math. Sociol.}}}
  \textbf{\bibinfo{volume}{1}}, \bibinfo{pages}{143--186}
  (\bibinfo{year}{1971}).

\bibitem{granovetter1978a}
\bibinfo{author}{Granovetter, M.}
\newblock \bibinfo{journal}{\bibinfo{title}{Threshold models of collective
  behavior}}.
\newblock {\emph{\JournalTitle{Am. J. Sociol.}}} \textbf{\bibinfo{volume}{83}},
  \bibinfo{pages}{1420--1443} (\bibinfo{year}{1978}).

\bibitem{granovetter1983a}
\bibinfo{author}{Granovetter, M.~S.} \& \bibinfo{author}{Soong, R.}
\newblock \bibinfo{journal}{\bibinfo{title}{Threshold models of diffusion and
  collective behavior}}.
\newblock {\emph{\JournalTitle{Journal of Mathematical Sociology}}}
  \textbf{\bibinfo{volume}{9}}, \bibinfo{pages}{165--179}
  (\bibinfo{year}{1983}).

\bibitem{granovetter1986a}
\bibinfo{author}{Granovetter, M.~S.} \& \bibinfo{author}{Soong, R.}
\newblock \bibinfo{journal}{\bibinfo{title}{Threshold models of interpersonal
  effects in consumer demand}}.
\newblock {\emph{\JournalTitle{J. Econ. Behav. Organ.}}}
  \textbf{\bibinfo{volume}{7}}, \bibinfo{pages}{83--99} (\bibinfo{year}{1986}).

\bibitem{granovetter1988a}
\bibinfo{author}{Granovetter, M.} \& \bibinfo{author}{Soong, R.}
\newblock \bibinfo{journal}{\bibinfo{title}{Threshold models of diversity:
  Chinese restaurants, residential segregation, and the spiral of silence}}.
\newblock {\emph{\JournalTitle{Sociological Methodology}}}
  \textbf{\bibinfo{volume}{18}}, \bibinfo{pages}{69--104}
  (\bibinfo{year}{1988}).

\bibitem{dodds2004a}
\bibinfo{author}{Dodds, P.~S.} \& \bibinfo{author}{Watts, D.~J.}
\newblock \bibinfo{journal}{\bibinfo{title}{Universal behavior in a generalized
  model of contagion}}.
\newblock {\emph{\JournalTitle{Phys. Rev. Lett.}}}
  \textbf{\bibinfo{volume}{92}}, \bibinfo{pages}{218701}
  (\bibinfo{year}{2004}).

\bibitem{dodds2005a}
\bibinfo{author}{Dodds, P.~S.} \& \bibinfo{author}{Watts, D.~J.}
\newblock \bibinfo{journal}{\bibinfo{title}{A generalized model of social and
  biological contagion}}.
\newblock {\emph{\JournalTitle{J. Theor. Biol.}}}
  \textbf{\bibinfo{volume}{232}}, \bibinfo{pages}{587--604},
  \doiprefix\url{doi:10.1016/j.jtbi.2004.09.006} (\bibinfo{year}{2005}).

\bibitem{guille2013information}
\bibinfo{author}{Guille, A.}, \bibinfo{author}{Hacid, H.},
  \bibinfo{author}{Favre, C.} \& \bibinfo{author}{Zighed, D.~A.}
\newblock \bibinfo{journal}{\bibinfo{title}{Information diffusion in online
  social networks: A survey}}.
\newblock {\emph{\JournalTitle{ACM Sigmod Record}}}
  \textbf{\bibinfo{volume}{42}}, \bibinfo{pages}{17--28}
  (\bibinfo{year}{2013}).

\bibitem{lerman2010information}
\bibinfo{author}{Lerman, K.} \& \bibinfo{author}{Ghosh, R.}
\newblock \bibinfo{title}{Information contagion: An empirical study of the
  spread of news on {Digg and Twitter} social networks}.
\newblock In \emph{\bibinfo{booktitle}{Proceedings of the international AAAI
  conference on web and social media}}, vol.~\bibinfo{volume}{4},
  \bibinfo{pages}{90--97} (\bibinfo{year}{2010}).

\bibitem{lu2015competition}
\bibinfo{author}{Lu, W.}, \bibinfo{author}{Chen, W.} \&
  \bibinfo{author}{Lakshmanan, L.~V.}
\newblock \bibinfo{journal}{\bibinfo{title}{From competition to
  complementarity: comparative influence diffusion and maximization}}.
\newblock {\emph{\JournalTitle{Proceedings of the VLDB Endowment}}}
  \textbf{\bibinfo{volume}{9}}, \bibinfo{pages}{60--71} (\bibinfo{year}{2015}).

\bibitem{myers2012clash}
\bibinfo{author}{Myers, S.~A.} \& \bibinfo{author}{Leskovec, J.}
\newblock \bibinfo{title}{Clash of the contagions: Cooperation and competition
  in information diffusion}.
\newblock In \emph{\bibinfo{booktitle}{2012 IEEE 12th international conference
  on data mining}}, \bibinfo{pages}{539--548} (\bibinfo{organization}{IEEE},
  \bibinfo{year}{2012}).

\bibitem{zarezade2017correlated}
\bibinfo{author}{Zarezade, A.}, \bibinfo{author}{Khodadadi, A.},
  \bibinfo{author}{Farajtabar, M.}, \bibinfo{author}{Rabiee, H.} \&
  \bibinfo{author}{Zha, H.}
\newblock \bibinfo{title}{Correlated cascades: Compete or cooperate}.
\newblock In \emph{\bibinfo{booktitle}{Proceedings of the AAAI Conference on
  Artificial Intelligence}}, vol.~\bibinfo{volume}{31} (\bibinfo{year}{2017}).

\bibitem{hawkes1971spectra}
\bibinfo{author}{Hawkes, A.~G.}
\newblock \bibinfo{journal}{\bibinfo{title}{Spectra of some self-exciting and
  mutually exciting point processes}}.
\newblock {\emph{\JournalTitle{Biometrika}}} \textbf{\bibinfo{volume}{58}},
  \bibinfo{pages}{83--90} (\bibinfo{year}{1971}).

\bibitem{weng2012competition}
\bibinfo{author}{Weng, L.}, \bibinfo{author}{Flammini, A.},
  \bibinfo{author}{Vespignani, A.} \& \bibinfo{author}{Menczer, F.}
\newblock \bibinfo{journal}{\bibinfo{title}{Competition among memes in a world
  with limited attention}}.
\newblock {\emph{\JournalTitle{Scientific Reports}}}
  \textbf{\bibinfo{volume}{2}}, \bibinfo{pages}{335} (\bibinfo{year}{2012}).

\bibitem{althouse2015enhancing}
\bibinfo{author}{Althouse, B.~M.} \emph{et~al.}
\newblock \bibinfo{journal}{\bibinfo{title}{Enhancing disease surveillance with
  novel data streams: challenges and opportunities}}.
\newblock {\emph{\JournalTitle{EPJ Data Science}}}
  \textbf{\bibinfo{volume}{4}}, \bibinfo{pages}{1--8} (\bibinfo{year}{2015}).

\bibitem{leskovec2009memetracking}
\bibinfo{author}{Leskovec, J.}, \bibinfo{author}{Backstrom, L.} \&
  \bibinfo{author}{Kleinberg, J.}
\newblock \bibinfo{title}{Meme-tracking and the dynamics of the news cycle}.
\newblock In \emph{\bibinfo{booktitle}{Proceedings of the 15th ACM SIGKDD
  international conference on Knowledge discovery and data mining}},
  \bibinfo{pages}{497--506} (\bibinfo{year}{2009}).

\bibitem{gruhl2004information}
\bibinfo{author}{Gruhl, D.}, \bibinfo{author}{Guha, R.},
  \bibinfo{author}{Liben-Nowell, D.} \& \bibinfo{author}{Tomkins, A.}
\newblock \bibinfo{title}{Information diffusion through blogspace}.
\newblock In \emph{\bibinfo{booktitle}{Proceedings of the 13th international
  conference on World Wide Web}}, \bibinfo{pages}{491--501}
  (\bibinfo{year}{2004}).

\bibitem{adar2005tracking}
\bibinfo{author}{Adar, E.} \& \bibinfo{author}{Adamic, L.~A.}
\newblock \bibinfo{title}{Tracking information epidemics in blogspace}.
\newblock In \emph{\bibinfo{booktitle}{The 2005 IEEE/WIC/ACM International
  Conference on Web Intelligence (WI'05)}}, \bibinfo{pages}{207--214}
  (\bibinfo{organization}{IEEE}, \bibinfo{year}{2005}).

\bibitem{kleinberg2002bursty}
\bibinfo{author}{Kleinberg, J.}
\newblock \bibinfo{title}{Bursty and hierarchical structure in streams}.
\newblock In \emph{\bibinfo{booktitle}{Proceedings of the eighth ACM SIGKDD
  international conference on Knowledge Discovery and Data Mining}},
  \bibinfo{pages}{91--101} (\bibinfo{year}{2002}).

\bibitem{mandl2004implementing}
\bibinfo{author}{Mandl, K.~D.} \emph{et~al.}
\newblock \bibinfo{journal}{\bibinfo{title}{Implementing syndromic
  surveillance: a practical guide informed by the early experience}}.
\newblock {\emph{\JournalTitle{Journal of the American Medical Informatics
  Association}}} \textbf{\bibinfo{volume}{11}}, \bibinfo{pages}{141--150}
  (\bibinfo{year}{2004}).

\bibitem{dubey2018memesequencer}
\bibinfo{author}{Dubey, A.}, \bibinfo{author}{Moro, E.},
  \bibinfo{author}{Cebrian, M.} \& \bibinfo{author}{Rahwan, I.}
\newblock \bibinfo{title}{Memesequencer: Sparse matching for embedding image
  macros}.
\newblock In \emph{\bibinfo{booktitle}{Proceedings of the 2018 World Wide Web
  Conference}}, \bibinfo{pages}{1225--1235} (\bibinfo{year}{2018}).

\bibitem{afridi2021multimodal}
\bibinfo{author}{Afridi, T.~H.}, \bibinfo{author}{Alam, A.},
  \bibinfo{author}{Khan, M.~N.}, \bibinfo{author}{Khan, J.} \&
  \bibinfo{author}{Lee, Y.-K.}
\newblock \bibinfo{title}{A multimodal memes classification: A survey and open
  research issues}.
\newblock In \emph{\bibinfo{booktitle}{Innovations in Smart Cities Applications
  Volume 4: The Proceedings of the 5th International Conference on Smart City
  Applications}}, \bibinfo{pages}{1451--1466}
  (\bibinfo{organization}{Springer}, \bibinfo{year}{2021}).

\bibitem{beskow2020evolution}
\bibinfo{author}{Beskow, D.~M.}, \bibinfo{author}{Kumar, S.} \&
  \bibinfo{author}{Carley, K.~M.}
\newblock \bibinfo{journal}{\bibinfo{title}{The evolution of political memes:
  Detecting and characterizing internet memes with multi-modal deep learning}}.
\newblock {\emph{\JournalTitle{Information Processing \& Management}}}
  \textbf{\bibinfo{volume}{57}}, \bibinfo{pages}{102170}
  (\bibinfo{year}{2020}).

\bibitem{ling2021dissecting}
\bibinfo{author}{Ling, C.} \emph{et~al.}
\newblock \bibinfo{journal}{\bibinfo{title}{Dissecting the meme magic:
  Understanding indicators of virality in image memes}}.
\newblock {\emph{\JournalTitle{Proceedings of the ACM on Human-Computer
  Interaction}}} \textbf{\bibinfo{volume}{5}}, \bibinfo{pages}{1--24}
  (\bibinfo{year}{2021}).

\bibitem{sartre1938a}
\bibinfo{author}{Sartre, J.-P.}
\newblock \emph{\bibinfo{title}{La Naus{\'e}e}}
  (\bibinfo{publisher}{{\'E}ditions Gallimard}, \bibinfo{year}{1938}).

\bibitem{ranke1967a}
\bibinfo{author}{Ranke, K.}
\newblock \bibinfo{journal}{\bibinfo{title}{Kategorienprobleme der
  {V}olksprosa}}.
\newblock {\emph{\JournalTitle{Fabula}}} \textbf{\bibinfo{volume}{9}},
  \bibinfo{pages}{4} (\bibinfo{year}{1967}).

\bibitem{fisher1984a}
\bibinfo{author}{Fisher, W.~R.}
\newblock \bibinfo{journal}{\bibinfo{title}{Narration as a human communication
  paradigm: {T}he case of public moral argument}}.
\newblock {\emph{\JournalTitle{Communications Monographs}}}
  \textbf{\bibinfo{volume}{51}}, \bibinfo{pages}{1--22} (\bibinfo{year}{1984}).

\bibitem{gould1994a}
\bibinfo{author}{Gould, S.~J.}
\newblock \bibinfo{journal}{\bibinfo{title}{So near and yet so far}}.
\newblock {\emph{\JournalTitle{The New York Review of Books}}}
  \textbf{\bibinfo{volume}{41}}, \bibinfo{pages}{26} (\bibinfo{year}{1994}).

\bibitem{ferrand2001a}
\bibinfo{author}{Ferrand, N.} \& \bibinfo{author}{Weil, M.}
\newblock \emph{\bibinfo{title}{Homo narrativus: {D}ix ans de recherche sur la
  topique romanesque}} (\bibinfo{publisher}{Universit{\'e} Paul-Val{\'e}ry de
  Montpellier}, \bibinfo{year}{2001}).

\bibitem{bruner2003a}
\bibinfo{author}{Bruner, J.~S.}
\newblock \emph{\bibinfo{title}{Making stories: Law, literature, life}}
  (\bibinfo{publisher}{Harvard University Press}, \bibinfo{year}{2003}).

\bibitem{boyd2010a}
\bibinfo{author}{Boyd, B.}
\newblock \emph{\bibinfo{title}{On the Origin of Stories: Evolution, Cognition,
  and Fiction}} (\bibinfo{publisher}{Belknap Press}, \bibinfo{year}{2010}).

\bibitem{moretti2013a}
\bibinfo{author}{Moretti, F.}
\newblock \emph{\bibinfo{title}{Distant Reading}} (\bibinfo{publisher}{Verso},
  \bibinfo{address}{New York}, \bibinfo{year}{2013}).

\bibitem{dodds2013b}
\bibinfo{author}{Dodds, P.~S.}
\newblock \bibinfo{title}{Homo {N}arrativus and the trouble with fame}.
\newblock \bibinfo{howpublished}{Nautilus Magazine} (\bibinfo{year}{2013}).

\bibitem{gottschall2013a}
\bibinfo{author}{Gottschall, J.}
\newblock \emph{\bibinfo{title}{The Storytelling Animal: How Stories Make Us
  Human}} (\bibinfo{publisher}{Mariner Books}, \bibinfo{year}{2013}).

\bibitem{maruna2015a}
\bibinfo{author}{Maruna, S.}
\newblock \bibinfo{title}{Foreword: Narrative criminology as the new
  mainstream}.
\newblock In \emph{\bibinfo{booktitle}{Narrative criminology}},
  \bibinfo{pages}{vii--x} (\bibinfo{publisher}{New York University Press},
  \bibinfo{year}{2015}).

\bibitem{shiller2017a}
\bibinfo{author}{Shiller, R.~J.}
\newblock \bibinfo{journal}{\bibinfo{title}{Narrative economics}}.
\newblock {\emph{\JournalTitle{American Economic Review}}}
  \textbf{\bibinfo{volume}{107}}, \bibinfo{pages}{967--1004}
  (\bibinfo{year}{2017}).

\bibitem{puchner2017a}
\bibinfo{author}{Puchner, M.}
\newblock \emph{\bibinfo{title}{The Written World: How Literature Shaped
  Civilization}} (\bibinfo{publisher}{Random}, \bibinfo{year}{2017}).

\bibitem{garland2022impact}
\bibinfo{author}{Garland, J.}, \bibinfo{author}{Ghazi-Zahedi, K.},
  \bibinfo{author}{Young, J.-G.}, \bibinfo{author}{H{\'e}bert-Dufresne, L.} \&
  \bibinfo{author}{Galesic, M.}
\newblock \bibinfo{journal}{\bibinfo{title}{Impact and dynamics of hate and
  counter speech online}}.
\newblock {\emph{\JournalTitle{EPJ Data Science}}}
  \textbf{\bibinfo{volume}{11}}, \bibinfo{pages}{3} (\bibinfo{year}{2022}).

\bibitem{labov1997narrative}
\bibinfo{author}{Labov, W.} \& \bibinfo{author}{Waletzky, J.}
\newblock \bibinfo{journal}{\bibinfo{title}{Narrative analysis: {Oral versions
  of personal experience.}}}
\newblock  (\bibinfo{year}{1997}).

\bibitem{barkun2013culture}
\bibinfo{author}{Barkun, M.}
\newblock \emph{\bibinfo{title}{A culture of conspiracy: Apocalyptic visions in
  contemporary {America}}}, vol.~\bibinfo{volume}{15} (\bibinfo{publisher}{Univ
  of California Press}, \bibinfo{year}{2013}).

\bibitem{holur2022side}
\bibinfo{author}{Holur, P.}, \bibinfo{author}{Wang, T.},
  \bibinfo{author}{Shahsavari, S.}, \bibinfo{author}{Tangherlini, T.} \&
  \bibinfo{author}{Roychowdhury, V.}
\newblock \bibinfo{journal}{\bibinfo{title}{Which side are you on?
  {Insider-Outsider classification in conspiracy-theoretic social media}}}.
\newblock {\emph{\JournalTitle{Proceedings of the 60th Annual Meeting of the
  Association for Computational Linguistics}}} \textbf{\bibinfo{volume}{1}},
  \bibinfo{pages}{4975--4987} (\bibinfo{year}{2022}).

\bibitem{holur2023my}
\bibinfo{author}{Holur, P.}, \bibinfo{author}{Chong, D.},
  \bibinfo{author}{Tangherlini, T.} \& \bibinfo{author}{Roychowdhury, V.}
\newblock \bibinfo{title}{My side, your side and the evidence: Discovering
  aligned actor groups and the narratives they weave}.
\newblock In \emph{\bibinfo{booktitle}{Proceedings of the 61st Annual Meeting
  of the Association for Computational Linguistics}},
  \bibinfo{pages}{8938--8952} (\bibinfo{year}{2023}).

\bibitem{tangherlini2016mommy}
\bibinfo{author}{Tangherlini, T.~R.} \emph{et~al.}
\newblock \bibinfo{journal}{\bibinfo{title}{{“Mommy blogs”} and the
  vaccination exemption narrative: results from a machine-learning approach for
  story aggregation on parenting social media sites}}.
\newblock {\emph{\JournalTitle{JMIR Public Health and Surveillance}}}
  \textbf{\bibinfo{volume}{2}}, \bibinfo{pages}{e6586} (\bibinfo{year}{2016}).

\bibitem{greimas1966elements}
\bibinfo{author}{Greimas, A.~J.}
\newblock \bibinfo{journal}{\bibinfo{title}{{\'E}l{\'e}ments pour une
  th{\'e}orie de l'interpr{\'e}tation du r{\'e}cit mythique}}.
\newblock {\emph{\JournalTitle{Communications}}} \textbf{\bibinfo{volume}{8}},
  \bibinfo{pages}{28--59} (\bibinfo{year}{1966}).

\bibitem{lehnert1981plot}
\bibinfo{author}{Lehnert, W.~G.}
\newblock \bibinfo{journal}{\bibinfo{title}{Plot units and narrative
  summarization}}.
\newblock {\emph{\JournalTitle{Cognitive Science}}}
  \textbf{\bibinfo{volume}{5}}, \bibinfo{pages}{293--331}
  (\bibinfo{year}{1981}).

\bibitem{tangherlini2020automated}
\bibinfo{author}{Tangherlini, T.~R.}, \bibinfo{author}{Shahsavari, S.},
  \bibinfo{author}{Shahbazi, B.}, \bibinfo{author}{Ebrahimzadeh, E.} \&
  \bibinfo{author}{Roychowdhury, V.}
\newblock \bibinfo{journal}{\bibinfo{title}{An automated pipeline for the
  discovery of conspiracy and conspiracy theory narrative frameworks:
  {Bridgegate, Pizzagate and storytelling on the web}}}.
\newblock {\emph{\JournalTitle{PloS One}}} \textbf{\bibinfo{volume}{15}},
  \bibinfo{pages}{e0233879} (\bibinfo{year}{2020}).

\bibitem{boole1854investigation}
\bibinfo{author}{Boole, G.}
\newblock \emph{\bibinfo{title}{An investigation of the laws of thought: on
  which are founded the mathematical theories of logic and probabilities}},
  vol.~\bibinfo{volume}{2} (\bibinfo{publisher}{Walton and Maberly},
  \bibinfo{year}{1854}).

\bibitem{eskerod1947lrets}
\bibinfo{author}{Esker{\"o}d, A.}
\newblock \emph{\bibinfo{title}{{\'L}rets {\"a}ring: etnologiska studier i
  sk{\"o}rdens och julens tro och sed}}.
\newblock Ph.D. thesis, \bibinfo{school}{Nordisk museet}
  (\bibinfo{year}{1947}).

\bibitem{anderson1923kaiser}
\bibinfo{author}{Anderson, W.}
\newblock \emph{\bibinfo{title}{Kaiser und Abt: Die Geschichte eines
  Schwanks}}, vol.~\bibinfo{volume}{42} (\bibinfo{publisher}{Suomalainen
  Tiedeakatemia, Academia Scientiarum Fennica}, \bibinfo{year}{1923}).

\bibitem{bauch2013social}
\bibinfo{author}{Bauch, C.~T.} \& \bibinfo{author}{Galvani, A.~P.}
\newblock \bibinfo{journal}{\bibinfo{title}{Social factors in epidemiology}}.
\newblock {\emph{\JournalTitle{Science}}} \textbf{\bibinfo{volume}{342}},
  \bibinfo{pages}{47--49} (\bibinfo{year}{2013}).

\bibitem{magarey2020information}
\bibinfo{author}{Magarey, R.~D.} \& \bibinfo{author}{Trexler, C.~M.}
\newblock \bibinfo{journal}{\bibinfo{title}{Information: A missing component in
  understanding and mitigating social epidemics}}.
\newblock {\emph{\JournalTitle{Humanities and Social Sciences Communications}}}
  \textbf{\bibinfo{volume}{7}}, \bibinfo{pages}{1--11} (\bibinfo{year}{2020}).

\bibitem{nettasinghe2024dynamics}
\bibinfo{author}{Nettasinghe, B.}, \bibinfo{author}{Percus, A.~G.} \&
  \bibinfo{author}{Lerman, K.}
\newblock \bibinfo{journal}{\bibinfo{title}{How out-group animosity can shape
  partisan divisions: A model of affective polarization}}.
\newblock {\emph{\JournalTitle{PNAS nexus}}} \bibinfo{pages}{pgaf082}
  (\bibinfo{year}{2025}).

\bibitem{nettasinghe2024group}
\bibinfo{author}{Nettasinghe, B.}, \bibinfo{author}{Rao, A.},
  \bibinfo{author}{Jiang, B.}, \bibinfo{author}{Percus, A.} \&
  \bibinfo{author}{Lerman, K.}
\newblock \bibinfo{title}{In-group love, out-group hate: A framework to measure
  affective polarization via contentious online discussions}.
\newblock In \emph{\bibinfo{booktitle}{Proceedings of The Web Conference}}
  (\bibinfo{year}{2025}).

\bibitem{ajzen1991theory}
\bibinfo{author}{Ajzen, I.}
\newblock \bibinfo{journal}{\bibinfo{title}{The theory of planned behavior}}.
\newblock {\emph{\JournalTitle{Organizational Behavior and Human Decision
  Processes}}}  (\bibinfo{year}{1991}).

\bibitem{monroe2008general}
\bibinfo{author}{Monroe, B.~M.} \& \bibinfo{author}{Read, S.~J.}
\newblock \bibinfo{journal}{\bibinfo{title}{A general connectionist model of
  attitude structure and change: the acs (attitudes as constraint satisfaction)
  model.}}
\newblock {\emph{\JournalTitle{Psychological Review}}}
  \textbf{\bibinfo{volume}{115}}, \bibinfo{pages}{733} (\bibinfo{year}{2008}).

\bibitem{olsson2024analogies}
\bibinfo{author}{Olsson, H.} \& \bibinfo{author}{Galesic, M.}
\newblock \bibinfo{journal}{\bibinfo{title}{Analogies for modeling belief
  dynamics}}.
\newblock {\emph{\JournalTitle{Trends in Cognitive Sciences}}}
  (\bibinfo{year}{2024}).

\bibitem{schmelz2021overcoming}
\bibinfo{author}{Schmelz, K.} \& \bibinfo{author}{Bowles, S.}
\newblock \bibinfo{journal}{\bibinfo{title}{Overcoming {COVID-19} vaccination
  resistance when alternative policies affect the dynamics of conformism,
  social norms, and crowding out}}.
\newblock {\emph{\JournalTitle{Proceedings of the National Academy of
  Sciences}}} \textbf{\bibinfo{volume}{118}}, \bibinfo{pages}{e2104912118}
  (\bibinfo{year}{2021}).

\bibitem{pew_abortion}
\bibinfo{author}{{Pew Research Center}}.
\newblock \bibinfo{title}{Public opinion on abortion} (\bibinfo{year}{2024}).

\bibitem{festinger1954theory}
\bibinfo{author}{Festinger, L.}
\newblock \bibinfo{journal}{\bibinfo{title}{A theory of social comparison
  processes}}.
\newblock {\emph{\JournalTitle{Human Relations}}} \textbf{\bibinfo{volume}{7}},
  \bibinfo{pages}{117--140} (\bibinfo{year}{1954}).

\bibitem{gallup_marriage}
\bibinfo{author}{Brenan, M.}
\newblock \bibinfo{title}{Same-sex relations, marriage still supported by most
  in {U.S.}}
\newblock \bibinfo{howpublished}{Gallup} (\bibinfo{year}{2024}).

\bibitem{ebner_extremism}
\bibinfo{author}{Ebner, J.}
\newblock \bibinfo{title}{From margins to mainstream: How extremism has
  conquered the political middle}.
\newblock \bibinfo{howpublished}{International Centre for Counter-Terrorism}
  (\bibinfo{year}{2023}).

\bibitem{pertwee2022epidemic}
\bibinfo{author}{Pertwee, E.}, \bibinfo{author}{Simas, C.} \&
  \bibinfo{author}{Larson, H.~J.}
\newblock \bibinfo{journal}{\bibinfo{title}{An epidemic of uncertainty: rumors,
  conspiracy theories and vaccine hesitancy}}.
\newblock {\emph{\JournalTitle{Nature Medicine}}}
  \textbf{\bibinfo{volume}{28}}, \bibinfo{pages}{456--459}
  (\bibinfo{year}{2022}).

\bibitem{solomon2000pride}
\bibinfo{author}{Solomon, S.}, \bibinfo{author}{Greenberg, J.} \&
  \bibinfo{author}{Pyszczynski, T.}
\newblock \bibinfo{journal}{\bibinfo{title}{Pride and prejudice: Fear of death
  and social behavior}}.
\newblock {\emph{\JournalTitle{Current Directions in Psychological Science}}}
  \textbf{\bibinfo{volume}{9}}, \bibinfo{pages}{200--204}
  (\bibinfo{year}{2000}).

\bibitem{castellano2009statistical}
\bibinfo{author}{Castellano, C.}, \bibinfo{author}{Fortunato, S.} \&
  \bibinfo{author}{Loreto, V.}
\newblock \bibinfo{journal}{\bibinfo{title}{Statistical physics of social
  dynamics}}.
\newblock {\emph{\JournalTitle{Reviews of Modern Physics}}}
  \textbf{\bibinfo{volume}{81}}, \bibinfo{pages}{591--646}
  (\bibinfo{year}{2009}).

\bibitem{flache2017models}
\bibinfo{author}{Flache, A.} \emph{et~al.}
\newblock \bibinfo{journal}{\bibinfo{title}{Models of social influence: Towards
  the next frontiers}}.
\newblock {\emph{\JournalTitle{JASSS-The Journal of Artificial Societies and
  Social Simulation}}} \textbf{\bibinfo{volume}{20}}, \bibinfo{pages}{2}
  (\bibinfo{year}{2017}).

\bibitem{redner2019reality}
\bibinfo{author}{Redner, S.}
\newblock \bibinfo{journal}{\bibinfo{title}{Reality-inspired voter models: A
  mini-review}}.
\newblock {\emph{\JournalTitle{Comptes Rendus Physique}}}
  \textbf{\bibinfo{volume}{20}}, \bibinfo{pages}{275--292}
  (\bibinfo{year}{2019}).

\bibitem{galesic2021integrating}
\bibinfo{author}{Galesic, M.}, \bibinfo{author}{Olsson, H.},
  \bibinfo{author}{Dalege, J.}, \bibinfo{author}{Van Der~Does, T.} \&
  \bibinfo{author}{Stein, D.~L.}
\newblock \bibinfo{journal}{\bibinfo{title}{Integrating social and cognitive
  aspects of belief dynamics: towards a unifying framework}}.
\newblock {\emph{\JournalTitle{Journal of the Royal Society Interface}}}
  \textbf{\bibinfo{volume}{18}}, \bibinfo{pages}{20200857}
  (\bibinfo{year}{2021}).

\bibitem{chater2006probabilistic}
\bibinfo{author}{Chater, N.}, \bibinfo{author}{Tenenbaum, J.~B.} \&
  \bibinfo{author}{Yuille, A.}
\newblock \bibinfo{journal}{\bibinfo{title}{Probabilistic models of cognition:
  Conceptual foundations}}.
\newblock {\emph{\JournalTitle{Trends in Cognitive Sciences}}}
  \textbf{\bibinfo{volume}{10}}, \bibinfo{pages}{287--291}
  (\bibinfo{year}{2006}).

\bibitem{bikhchandani2010theory}
\bibinfo{author}{Bikhchandani, S.}, \bibinfo{author}{Hirshleifer, D.} \&
  \bibinfo{author}{Welch, I.}
\newblock \bibinfo{title}{A theory of fads, fashion, custom and cultural change
  as informational cascades}.
\newblock \bibinfo{type}{Tech. Rep.}, \bibinfo{institution}{David K. Levine}
  (\bibinfo{year}{2010}).

\bibitem{van2011logical}
\bibinfo{author}{Van~Benthem, J.}
\newblock \emph{\bibinfo{title}{Logical dynamics of information and
  interaction}} (\bibinfo{publisher}{Cambridge University Press},
  \bibinfo{year}{2011}).

\bibitem{friedkin1990social}
\bibinfo{author}{Friedkin, N.~E.} \& \bibinfo{author}{Johnsen, E.~C.}
\newblock \bibinfo{journal}{\bibinfo{title}{Social influence and opinions}}.
\newblock {\emph{\JournalTitle{Journal of Mathematical Sociology}}}
  \textbf{\bibinfo{volume}{15}}, \bibinfo{pages}{193--206}
  (\bibinfo{year}{1990}).

\bibitem{einhorn1975unit}
\bibinfo{author}{Einhorn, H.~J.} \& \bibinfo{author}{Hogarth, R.~M.}
\newblock \bibinfo{journal}{\bibinfo{title}{Unit weighting schemes for decision
  making}}.
\newblock {\emph{\JournalTitle{Organizational Behavior and Human Performance}}}
  \textbf{\bibinfo{volume}{13}}, \bibinfo{pages}{171--192}
  (\bibinfo{year}{1975}).

\bibitem{fic2024catalysing}
\bibinfo{author}{Fic, M.} \& \bibinfo{author}{Gokhale, C.~S.}
\newblock \bibinfo{journal}{\bibinfo{title}{Catalysing cooperation: the power
  of collective beliefs in structured populations}}.
\newblock {\emph{\JournalTitle{npj Complexity}}} \textbf{\bibinfo{volume}{1}},
  \bibinfo{pages}{6} (\bibinfo{year}{2024}).

\bibitem{gigerenzer1996reasoning}
\bibinfo{author}{Gigerenzer, G.} \& \bibinfo{author}{Goldstein, D.~G.}
\newblock \bibinfo{journal}{\bibinfo{title}{Reasoning the fast and frugal way:
  models of bounded rationality.}}
\newblock {\emph{\JournalTitle{Psychological Review}}}
  \textbf{\bibinfo{volume}{103}}, \bibinfo{pages}{650} (\bibinfo{year}{1996}).

\bibitem{converse1964}
\bibinfo{author}{Converse, P.~E.}
\newblock \bibinfo{title}{The nature of belief systems in mass publics}.
\newblock In \bibinfo{editor}{Apter, D.~E.} (ed.)
  \emph{\bibinfo{booktitle}{Ideology and Its Disconte}},
  \bibinfo{pages}{206--261} (\bibinfo{publisher}{The Free Press},
  \bibinfo{address}{New York, NY, USA}, \bibinfo{year}{1964}).

\bibitem{dalege2016toward}
\bibinfo{author}{Dalege, J.} \emph{et~al.}
\newblock \bibinfo{journal}{\bibinfo{title}{Toward a formalized account of
  attitudes: The {Causal Attitude Network (CAN)} model.}}
\newblock {\emph{\JournalTitle{Psychological Review}}}
  \textbf{\bibinfo{volume}{123}}, \bibinfo{pages}{2} (\bibinfo{year}{2016}).

\bibitem{boutyline2017belief}
\bibinfo{author}{Boutyline, A.} \& \bibinfo{author}{Vaisey, S.}
\newblock \bibinfo{journal}{\bibinfo{title}{Belief network analysis: A
  relational approach to understanding the structure of attitudes}}.
\newblock {\emph{\JournalTitle{American Journal of Sociology}}}
  \textbf{\bibinfo{volume}{122}}, \bibinfo{pages}{1371--1447}
  (\bibinfo{year}{2017}).

\bibitem{brandt2019central}
\bibinfo{author}{Brandt, M.~J.}, \bibinfo{author}{Sibley, C.~G.} \&
  \bibinfo{author}{Osborne, D.}
\newblock \bibinfo{journal}{\bibinfo{title}{What is central to political belief
  system networks?}}
\newblock {\emph{\JournalTitle{Personality and Social Psychology Bulletin}}}
  \textbf{\bibinfo{volume}{45}}, \bibinfo{pages}{1352--1364}
  (\bibinfo{year}{2019}).

\bibitem{vlasceanu2024network}
\bibinfo{author}{Vlasceanu, M.}, \bibinfo{author}{Dyckovsky, A.~M.} \&
  \bibinfo{author}{Coman, A.}
\newblock \bibinfo{journal}{\bibinfo{title}{A network approach to investigate
  the dynamics of individual and collective beliefs: Advances and applications
  of the bending model}}.
\newblock {\emph{\JournalTitle{Perspectives on Psychological Science}}}
  \textbf{\bibinfo{volume}{19}}, \bibinfo{pages}{444--453}
  (\bibinfo{year}{2024}).

\bibitem{rodriguez2016collective}
\bibinfo{author}{Rodriguez, N.}, \bibinfo{author}{Bollen, J.} \&
  \bibinfo{author}{Ahn, Y.-Y.}
\newblock \bibinfo{journal}{\bibinfo{title}{Collective dynamics of belief
  evolution under cognitive coherence and social conformity}}.
\newblock {\emph{\JournalTitle{PLoS One}}} \textbf{\bibinfo{volume}{11}},
  \bibinfo{pages}{e0165910} (\bibinfo{year}{2016}).

\bibitem{brandt2023inter}
\bibinfo{author}{Brandt, M.~J.} \& \bibinfo{author}{Vallabha, S.}
\newblock \bibinfo{journal}{\bibinfo{title}{Inter-attitude centrality does not
  appear to reduce persuasion for political attitudes}}.
\newblock {\emph{\JournalTitle{European Journal of Social Psychology}}}
  \textbf{\bibinfo{volume}{53}}, \bibinfo{pages}{1342--1358}
  (\bibinfo{year}{2023}).

\bibitem{turner2022belief}
\bibinfo{author}{Turner-Zwinkels, F.~M.} \& \bibinfo{author}{Brandt, M.~J.}
\newblock \bibinfo{journal}{\bibinfo{title}{Belief system networks can be used
  to predict where to expect dynamic constraint}}.
\newblock {\emph{\JournalTitle{Journal of Experimental Social Psychology}}}
  \textbf{\bibinfo{volume}{100}}, \bibinfo{pages}{104279}
  (\bibinfo{year}{2022}).

\bibitem{dalege2024networks}
\bibinfo{author}{Dalege, J.}, \bibinfo{author}{Galesic, M.} \&
  \bibinfo{author}{Olsson, H.}
\newblock \bibinfo{journal}{\bibinfo{title}{Networks of beliefs: An integrative
  theory of individual-and social-level belief dynamics.}}
\newblock {\emph{\JournalTitle{Psychological Review}}}  (\bibinfo{year}{2024}).

\bibitem{aiyappa2024emergence}
\bibinfo{author}{Aiyappa, R.}, \bibinfo{author}{Flammini, A.} \&
  \bibinfo{author}{Ahn, Y.-Y.}
\newblock \bibinfo{journal}{\bibinfo{title}{Emergence of simple and complex
  contagion dynamics from weighted belief networks}}.
\newblock {\emph{\JournalTitle{Science Advances}}}
  \textbf{\bibinfo{volume}{10}}, \bibinfo{pages}{eadh4439}
  (\bibinfo{year}{2024}).

\bibitem{maghool2019coevolution}
\bibinfo{author}{Maghool, S.}, \bibinfo{author}{Maleki-Jirsaraei, N.} \&
  \bibinfo{author}{Cremonini, M.}
\newblock \bibinfo{journal}{\bibinfo{title}{The coevolution of contagion and
  behavior with increasing and decreasing awareness}}.
\newblock {\emph{\JournalTitle{PloS One}}} \textbf{\bibinfo{volume}{14}},
  \bibinfo{pages}{e0225447} (\bibinfo{year}{2019}).

\bibitem{ogara2024}
\bibinfo{author}{O’Gara, D.}, \bibinfo{author}{Kasman, M.},
  \bibinfo{author}{H{\'e}bert-Dufresne, L.} \& \bibinfo{author}{Hammond, R.~A.}
\newblock \bibinfo{journal}{\bibinfo{title}{Adaptive behaviour during
  epidemics: a social risk appraisal approach to modelling dynamics}}.
\newblock {\emph{\JournalTitle{Journal of the Royal Society Interface}}}
  \textbf{\bibinfo{volume}{22}}, \bibinfo{pages}{20240363}
  (\bibinfo{year}{2025}).

\bibitem{stanoev2014modeling}
\bibinfo{author}{Stanoev, A.}, \bibinfo{author}{Trpevski, D.} \&
  \bibinfo{author}{Kocarev, L.}
\newblock \bibinfo{journal}{\bibinfo{title}{Modeling the spread of multiple
  concurrent contagions on networks}}.
\newblock {\emph{\JournalTitle{PloS One}}} \textbf{\bibinfo{volume}{9}},
  \bibinfo{pages}{e95669} (\bibinfo{year}{2014}).

\bibitem{shrestha2011statistical}
\bibinfo{author}{Shrestha, S.}, \bibinfo{author}{King, A.~A.} \&
  \bibinfo{author}{Rohani, P.}
\newblock \bibinfo{journal}{\bibinfo{title}{Statistical inference for
  multi-pathogen systems}}.
\newblock {\emph{\JournalTitle{PLoS Computational Biology}}}
  \textbf{\bibinfo{volume}{7}}, \bibinfo{pages}{e1002135}
  (\bibinfo{year}{2011}).

\bibitem{colgate2023network}
\bibinfo{author}{Colgate, E.~R.} \emph{et~al.}
\newblock \bibinfo{journal}{\bibinfo{title}{Network analysis of patterns and
  relevance of enteric pathogen co-infections among infants in a
  diarrhea-endemic setting}}.
\newblock {\emph{\JournalTitle{PLOS Computational Biology}}}
  \textbf{\bibinfo{volume}{19}}, \bibinfo{pages}{e1011624}
  (\bibinfo{year}{2023}).

\bibitem{fountain2019endemic}
\bibinfo{author}{Fountain-Jones, N.~M.} \emph{et~al.}
\newblock \bibinfo{journal}{\bibinfo{title}{Endemic infection can shape
  exposure to novel pathogens: {Pathogen co-occurrence networks in the
  Serengeti lions}}}.
\newblock {\emph{\JournalTitle{Ecology Letters}}}
  \textbf{\bibinfo{volume}{22}}, \bibinfo{pages}{904--913}
  (\bibinfo{year}{2019}).

\bibitem{gul2024stance}
\bibinfo{author}{G{\"u}l, I.}, \bibinfo{author}{Lebret, R.} \&
  \bibinfo{author}{Aberer, K.}
\newblock \bibinfo{journal}{\bibinfo{title}{Stance detection on social media
  with fine-tuned large language models}}.
\newblock {\emph{\JournalTitle{arXiv preprint arXiv:2404.12171}}}
  (\bibinfo{year}{2024}).

\bibitem{blanchet2020co}
\bibinfo{author}{Blanchet, F.~G.}, \bibinfo{author}{Cazelles, K.} \&
  \bibinfo{author}{Gravel, D.}
\newblock \bibinfo{journal}{\bibinfo{title}{Co-occurrence is not evidence of
  ecological interactions}}.
\newblock {\emph{\JournalTitle{Ecology Letters}}}
  \textbf{\bibinfo{volume}{23}}, \bibinfo{pages}{1050--1063}
  (\bibinfo{year}{2020}).

\bibitem{powell2024systematic}
\bibinfo{author}{Powell-Romero, F.}, \bibinfo{author}{Wells, K.} \&
  \bibinfo{author}{Clark, N.~J.}
\newblock \bibinfo{journal}{\bibinfo{title}{A systematic review and guide for
  using multi-response statistical models in co-infection research}}.
\newblock {\emph{\JournalTitle{Royal Society Open Science}}}
  \textbf{\bibinfo{volume}{11}}, \bibinfo{pages}{231589}
  (\bibinfo{year}{2024}).

\bibitem{chen2019persistent}
\bibinfo{author}{Chen, L.}
\newblock \bibinfo{journal}{\bibinfo{title}{Persistent spatial patterns of
  interacting contagions}}.
\newblock {\emph{\JournalTitle{Physical Review E}}}
  \textbf{\bibinfo{volume}{99}}, \bibinfo{pages}{022308}
  (\bibinfo{year}{2019}).

\bibitem{waring2023operationalizing}
\bibinfo{author}{Waring, T.~M.} \emph{et~al.}
\newblock \bibinfo{journal}{\bibinfo{title}{Operationalizing cultural
  adaptation to climate change: Contemporary examples from {United States
  agriculture}}}.
\newblock {\emph{\JournalTitle{Proceedings of the Royal Society B}}}
  \textbf{\bibinfo{volume}{378}}, \bibinfo{pages}{20220397}
  (\bibinfo{year}{2023}).

\bibitem{griffiths2011nature}
\bibinfo{author}{Griffiths, E.~C.}, \bibinfo{author}{Pedersen, A.~B.},
  \bibinfo{author}{Fenton, A.} \& \bibinfo{author}{Petchey, O.~L.}
\newblock \bibinfo{journal}{\bibinfo{title}{The nature and consequences of
  coinfection in humans}}.
\newblock {\emph{\JournalTitle{Journal of Infection}}}
  \textbf{\bibinfo{volume}{63}}, \bibinfo{pages}{200--206}
  (\bibinfo{year}{2011}).

\bibitem{su2016understanding}
\bibinfo{author}{Su, Y.}, \bibinfo{author}{Zhang, X.}, \bibinfo{author}{Liu,
  L.}, \bibinfo{author}{Song, S.} \& \bibinfo{author}{Fang, B.}
\newblock \bibinfo{journal}{\bibinfo{title}{Understanding information
  interactions in diffusion: an evolutionary game-theoretic perspective}}.
\newblock {\emph{\JournalTitle{Frontiers of Computer Science}}}
  \textbf{\bibinfo{volume}{10}}, \bibinfo{pages}{518--531}
  (\bibinfo{year}{2016}).

\bibitem{ghanbarnejad2022emergence}
\bibinfo{author}{Ghanbarnejad, F.}, \bibinfo{author}{Seegers, K.},
  \bibinfo{author}{Cardillo, A.} \& \bibinfo{author}{H{\"o}vel, P.}
\newblock \bibinfo{journal}{\bibinfo{title}{Emergence of synergistic and
  competitive pathogens in a coevolutionary spreading model}}.
\newblock {\emph{\JournalTitle{Physical Review E}}}
  \textbf{\bibinfo{volume}{105}}, \bibinfo{pages}{034308}
  (\bibinfo{year}{2022}).

\bibitem{khazaee2022effects}
\bibinfo{author}{Khazaee, A.} \& \bibinfo{author}{Ghanbarnejad, F.}
\newblock \bibinfo{journal}{\bibinfo{title}{Effects of measures on phase
  transitions in two cooperative susceptible-infectious-recovered dynamics}}.
\newblock {\emph{\JournalTitle{Physical Review E}}}
  \textbf{\bibinfo{volume}{105}}, \bibinfo{pages}{034311}
  (\bibinfo{year}{2022}).

\bibitem{levins1969some}
\bibinfo{author}{Levins, R.}
\newblock \bibinfo{journal}{\bibinfo{title}{Some demographic and genetic
  consequences of environmental heterogeneity for biological control}}.
\newblock {\emph{\JournalTitle{Bulletin of the Entomological Society of
  America}}} \textbf{\bibinfo{volume}{15}}, \bibinfo{pages}{237--240}
  (\bibinfo{year}{1969}).

\bibitem{hanski1983coexistence}
\bibinfo{author}{Hanski, I.}
\newblock \bibinfo{journal}{\bibinfo{title}{Coexistence of competitors in
  patchy environment}}.
\newblock {\emph{\JournalTitle{Ecology}}} \textbf{\bibinfo{volume}{64}},
  \bibinfo{pages}{493--500} (\bibinfo{year}{1983}).

\bibitem{cazelles2016integration}
\bibinfo{author}{Cazelles, K.}, \bibinfo{author}{Mouquet, N.},
  \bibinfo{author}{Mouillot, D.} \& \bibinfo{author}{Gravel, D.}
\newblock \bibinfo{journal}{\bibinfo{title}{On the integration of biotic
  interaction and environmental constraints at the biogeographical scale}}.
\newblock {\emph{\JournalTitle{Ecography}}} \textbf{\bibinfo{volume}{39}},
  \bibinfo{pages}{921--931} (\bibinfo{year}{2016}).

\bibitem{hassell2021towards}
\bibinfo{author}{Hassell, J.~M.} \emph{et~al.}
\newblock \bibinfo{journal}{\bibinfo{title}{Towards an ecosystem model of
  infectious disease}}.
\newblock {\emph{\JournalTitle{Nature Ecology \& Evolution}}}
  \textbf{\bibinfo{volume}{5}}, \bibinfo{pages}{907--918}
  (\bibinfo{year}{2021}).

\bibitem{vila2021viewing}
\bibinfo{author}{Vil{\`a}, M.} \emph{et~al.}
\newblock \bibinfo{journal}{\bibinfo{title}{Viewing emerging human infectious
  epidemics through the lens of invasion biology}}.
\newblock {\emph{\JournalTitle{BioScience}}} \textbf{\bibinfo{volume}{71}},
  \bibinfo{pages}{722--740} (\bibinfo{year}{2021}).

\bibitem{lasky2020processes}
\bibinfo{author}{Lasky, J.~R.}, \bibinfo{author}{Hooten, M.~B.} \&
  \bibinfo{author}{Adler, P.~B.}
\newblock \bibinfo{journal}{\bibinfo{title}{What processes must we understand
  to forecast regional-scale population dynamics?}}
\newblock {\emph{\JournalTitle{Proceedings of the Royal Society B}}}
  \textbf{\bibinfo{volume}{287}}, \bibinfo{pages}{20202219}
  (\bibinfo{year}{2020}).

\end{thebibliography}
\end{document}